\pdfoutput=1

\documentclass[twocolumn,showpacs,aps,prd,superscriptaddress]{revtex4-2}

\usepackage[title]{appendix}
\usepackage{hyperref} 
\usepackage{graphicx}
\usepackage{dcolumn}
\usepackage{amsmath}
\usepackage{amssymb}
\usepackage{epsfig}
\usepackage{hhline}
\usepackage{soul}
\usepackage{tabularx,pifont,multirow}
\usepackage{subfigure}
\usepackage{bm} 
\usepackage{xcolor}
\usepackage{colordvi}
\usepackage[usenames,dvipsnames]{xcolor}
\definecolor{taplum}{rgb}{0.67843, 0.49804, 0.65882}
\RequirePackage{xspace}
\usepackage{grffile}

\usepackage{booktabs}
\usepackage{makecell}

\usepackage{lineno} 

\usepackage[absolute,overlay]{textpos} 

\usepackage{lmodern}  

\long\def\inst#1{\par\nobreak\kern 4pt\nobreak
  {\it #1}\par\vskip 10pt plus 3pt minus 3pt}

\usepackage{xspace} 
\usepackage{upgreek}

\def\lhcb {\mbox{LHCb}\xspace}

\def\MagUp {\mbox{\em Mag\kern -0.05em Up}\xspace}

 \def\PLambda     {\ensuremath{\Lambda}\xspace}                 
 \def\Pmu         {\ensuremath{\mu}\xspace}

 \def\Ppi         {\ensuremath{\pi}\xspace}

 \def\Ppsi        {\ensuremath{\psi}\xspace}                 
                  
 \def\PB      {\ensuremath{B}\xspace}                 
                  
 \def\PD      {\ensuremath{D}\xspace}

 \def\PJ      {\ensuremath{J}\xspace}                 
 \def\PK      {\ensuremath{K}\xspace}

 \def\Pb      {\ensuremath{b}\xspace}                 
 \def\Pc      {\ensuremath{c}\xspace}

 \def\Pi      {\ensuremath{i}\xspace}

\def\mup        {{\ensuremath{\Pmu^+}}\xspace}
\def\mun        {{\ensuremath{\Pmu^-}}\xspace} 

\def\cquark    {{\ensuremath{\Pc}}\xspace}

\def\bquark    {{\ensuremath{\Pb}}\xspace}

\def\pion   {{\ensuremath{\Ppi}}\xspace}

\def\pip    {{\ensuremath{\pion^+}}\xspace}
\def\pim    {{\ensuremath{\pion^-}}\xspace}

\def\kaon    {{\ensuremath{\PK}}\xspace}
  \def\Kbar    {{\kern 0.2em\overline{\kern -0.2em \PK}{}}\xspace}

\def\KorKbar    {\kern 0.18em\optbar{\kern -0.18em K}{}\xspace}

\def\KS      {{\ensuremath{\kaon^0_{\mathrm{ \scriptscriptstyle S}}}}\xspace}

  \def\Dbar    {{\kern 0.2em\overline{\kern -0.2em \PD}{}}\xspace}

\def\DorDbar    {\kern 0.18em\optbar{\kern -0.18em D}{}\xspace}

\def\B       {{\ensuremath{\PB}}\xspace}
\def\Bbar    {{\ensuremath{\kern 0.18em\overline{\kern -0.18em \PB}{}}}\xspace}

\def\BorBbar    {\kern 0.18em\optbar{\kern -0.18em B}{}\xspace}

\def\Bd      {{\ensuremath{\B^0}}\xspace}

\def\jpsi     {{\ensuremath{{\PJ\mskip -3mu/\mskip -2mu\Ppsi\mskip 2mu}}}\xspace}

  \def\Y#1S{\ensuremath{\PUpsilon{(#1S)}}\xspace}

\def\Lz          {{\ensuremath{\PLambda}}\xspace}
\def\Lbar        {{\ensuremath{\kern 0.1em\overline{\kern -0.1em\PLambda}}}\xspace}
\def\LorLbar    {\kern 0.18em\optbar{\kern -0.18em \PLambda}{}\xspace}

\def\Lb      {{\ensuremath{\Lz^0_\bquark}}\xspace}

\def\to                 {\ensuremath{\rightarrow}\xspace}

\def\AT#1     {\ensuremath{A_{\mathrm{T}}^{#1}}\xspace}           

\def\C#1      {\ensuremath{\mathcal{C}_{#1}}\xspace}                       
\def\Cp#1     {\ensuremath{\mathcal{C}_{#1}^{'}}\xspace}                    
\def\Ceff#1   {\ensuremath{\mathcal{C}_{#1}^{\mathrm{(eff)}}}\xspace}        
\def\Cpeff#1  {\ensuremath{\mathcal{C}_{#1}^{'\mathrm{(eff)}}}\xspace}       
\def\Ope#1    {\ensuremath{\mathcal{O}_{#1}}\xspace}                       
\def\Opep#1   {\ensuremath{\mathcal{O}_{#1}^{'}}\xspace}                    

\newcommand{\tev}{\ensuremath{\mathrm{\,Te\kern -0.1em V}}\xspace}
\newcommand{\gev}{\ensuremath{\mathrm{\,Ge\kern -0.1em V}}\xspace}
\newcommand{\mev}{\ensuremath{\mathrm{\,Me\kern -0.1em V}}\xspace}
\newcommand{\kev}{\ensuremath{\mathrm{\,ke\kern -0.1em V}}\xspace}
\newcommand{\ev}{\ensuremath{\mathrm{\,e\kern -0.1em V}}\xspace}
\newcommand{\gevc}{\ensuremath{{\mathrm{\,Ge\kern -0.1em V\!/}c}}\xspace}
\newcommand{\mevc}{\ensuremath{{\mathrm{\,Me\kern -0.1em V\!/}c}}\xspace}
\newcommand{\gevcc}{\ensuremath{{\mathrm{\,Ge\kern -0.1em V\!/}c^2}}\xspace}
\newcommand{\gevgevcccc}{\ensuremath{{\mathrm{\,Ge\kern -0.1em V^2\!/}c^4}}\xspace}
\newcommand{\mevcc}{\ensuremath{{\mathrm{\,Me\kern -0.1em V\!/}c^2}}\xspace}

\def\m    {\ensuremath{\mathrm{ \,m}}\xspace}

\def\cm   {\ensuremath{\mathrm{ \,cm}}\xspace}

\def\mm   {\ensuremath{\mathrm{ \,mm}}\xspace}

\def\mum  {\ensuremath{{\,\upmu\mathrm{m}}}\xspace}

\def\gsim{{~\raise.15em\hbox{$>$}\kern-.85em
          \lower.35em\hbox{$\sim$}~}\xspace}
\def\lsim{{~\raise.15em\hbox{$<$}\kern-.85em
          \lower.35em\hbox{$\sim$}~}\xspace}

\def\evtgen     {\mbox{\textsc{EvtGen}}\xspace}

\def\geant      {\mbox{\textsc{Geant4}}\xspace}

\def\photos     {\mbox{\textsc{Photos++}}\xspace}

\def\pythia     {\mbox{\textsc{Pythia}}\xspace}

\def\cpp        {\mbox{\textsc{C\raisebox{0.1em}{{\footnotesize{++}}}}}\xspace}

\def\tell1  {TELL1\xspace}
\def\ukl1   {UKL1\xspace}

\newcommand{\eg}{\mbox{\itshape e.g.}\xspace}
\newcommand{\ie}{\mbox{\itshape i.e.}\xspace}

\def\pr          {{\ensuremath{p}}\xspace}

\newcommand{\tevc}{\ensuremath{{\mathrm{\,Te\kern -0.1em V\!/}c}}\xspace}
\newcommand{\tevtevcccc}{\ensuremath{{\mathrm{\,Te\kern -0.1em V^2\!/}c^4}}\xspace}
\newcommand{\gevgevcc}{\ensuremath{{\mathrm{\,Ge\kern -0.1em V^2\!/}c^2}}\xspace}
\newcommand{\tevtevcc}{\ensuremath{{\mathrm{\,Te\kern -0.1em V^2\!/}c^2}}\xspace}

\def\C                {{\ensuremath{C}}\xspace}

\def\trkbare   { {\mathrm tr} }

\def\meas      { {\mathrm (m)} }

\newcommand{\forwardtreefitter}{\textsc{ForwardTreeFitter}\xspace}
\newcommand{\ftransport}{\textsc{FTransport}\xspace}
\newcommand{\fdoca}{\textsc{FDoca}\xspace}
\newcommand{\fvertexer}{\textsc{FVertexer}\xspace}
\newcommand{\ftreefitter}{\textsc{FTreeFitter}\xspace}
\newcommand{\ftf}{\textsc{fTf}\xspace}

\newcommand{\particlevertexfitter}{\textsc{ParticleVertexFitter}\xspace}
\newcommand{\pvf}{\textsc{PVF}\xspace}

\newcommand{\decaytreefitter}{\textsc{DecayTreeFitter}\xspace}
\newcommand{\dtf}{\textsc{DTF}\xspace}

\long\def\inst#1{\par\nobreak\kern 4pt\nobreak
  {\it #1}\par\vskip 10pt plus 3pt minus 3pt}

\begin{document}



\begin{flushright}
\begin{tabular}{r}
\normalsize CERN-LHCb-RD-2026-007 \\
\normalsize CERN-LHCb-DP-2026-002
\end{tabular}
\end{flushright}

\title{
{\large \bf \boldmath
Forward decay chain reconstruction with displaced \\ vertices in intense magnetic fields
}
}

\author{S. Jaimes Elles}
\affiliation{IFIC, Universitat de Val\`encia-CSIC, Valencia, Spain}
\affiliation{European Organization for Nuclear Research (CERN), Geneva, Switzerland}
\author{S. Libralon}
\affiliation{IFIC, Universitat de Val\`encia-CSIC, Valencia, Spain}
\author{F.~Martinez Vidal}
\email{fernando.martinez.vidal@cern.ch}
\affiliation{IFIC, Universitat de Val\`encia-CSIC, Valencia, Spain}
\author{N.~Neri}
\affiliation{INFN Sezione di Milano, Milan, Italy}
\affiliation{Università degli Studi di Milano, Milan, Italy}
\author{M.~Rebollo de Miguel}
\affiliation{IFIC, Universitat de Val\`encia-CSIC, Valencia, Spain}
\author{I.~Sanderswood}
\email{izaac.sanderswood@cern.ch}
\affiliation{IFIC, Universitat de Val\`encia-CSIC, Valencia, Spain}
\author{V.~Svintozelskyi}
\affiliation{IFIC, Universitat de Val\`encia-CSIC, Valencia, Spain}
\affiliation{European Organization for Nuclear Research (CERN), Geneva, Switzerland}
\author{G.~Tonani}
\affiliation{INFN Sezione di Milano, Milan, Italy}
\affiliation{Università degli Studi di Milano, Milan, Italy}
\author{J.~Velilla Serna}
\affiliation{IFIC, Universitat de Val\`encia-CSIC, Valencia, Spain}
\author{Z.~Wang}
\affiliation{INFN Sezione di Milano, Milan, Italy}
\affiliation{Università degli Studi di Milano, Milan, Italy}

\date{\today}

\begin{abstract}
  Precise reconstruction of particle decay chains is essential for a broad range of analyses in particle physics experiments, particularly those targeting flavour physics and spectroscopy. We present a standalone, modular and extensible decay tree reconstruction framework designed to handle complicated topologies, accounting for tightly constrained particle decays and long-distance trajectory extrapolations through regions of intense and inhomogeneous magnetic field, as well as limited momentum resolution. Implemented in \cpp, the framework is optimised for configurability and performance. Developed primarily within CERN's LHCb experiment, its focus is on decay chains with vertices located several metres from the LHC proton-proton interaction point, where the dipole magnetic field is maximal and non-uniform. Even so, it remains flexible: it can accommodate these special cases as well as the approximations commonly used in standard decay tree fitting tools, and the geometries of general-purpose collider detectors.
Benchmark studies using simulated data demonstrate significant improvements in reconstruction efficiency, invariant mass resolution, and vertex position resolution with respect to existing tools, opening new opportunities for precision measurements and searches involving highly displaced decays.
\end{abstract}
    
\maketitle

\begin{textblock*}{0.90\textwidth}(0.05\textwidth,1.10\textheight)
\centering
\footnotesize
Submitted to \textit{Eur. Phys. J. C - Computing, Software and Data Science}
\end{textblock*}

\section{Introduction}
\label{sec:introduction}

The accurate reconstruction of particle decay chains is fundamental for many precise measurements of particle properties and interactions. This is especially critical in flavour physics, where complex decay topologies involving multiple final-state particles, and intermediate long- and short-lived particles and resonances are common. Standard implementations of vertex and kinematic fitting algorithms depend on assumptions such as 
the adequate description of particle trajectories by local linear approximations at the decay point,
which work well for most conventional experimental cases. However, these assumptions break down when dealing with hierarchical decay trees and constrained systems in certain experimental scenarios, such as when the decays occur in regions of intense and inhomogeneous magnetic field, the charged particles traverse large detector volumes between its production and first measurement surface, and there is limited momentum resolution. Decay chain reconstruction with existing tools is therefore limited in efficiency and resolution under these conditions, motivating the development of a more general and flexible solution. 

The proposed framework was primarily developed for the \lhcb experiment~\cite{LHCb:2023hlw} to resolve these challenging scenarios. The performance of existing tools under these experimental conditions has been studied in detail in~\cite{LHCb:2022kwc}.
The framework remains sufficiently flexible to accommodate both the approximations commonly employed in standard decay tree fitting tools and the $4\pi$ geometry of general-purpose collider detectors. Its development was motivated by the reconstruction of \bquark- and \cquark-hadron decays containing long-lived particles that decay between 2.5 and 7.5\,m from the LHC interaction point (IP)~\cite{LHCb:2022kwc}, where the \lhcb dipolar magnetic field is maximal and non-uniform. There, particle decays are reconstructed using tracks with a limited momentum resolution.

The algorithms are implemented as a C++ framework named \forwardtreefitter, publicly available through CERN~\cite{ForwardTreeFitter2025:website} and distributed under the Mozilla Public License 2.0 (MPL-2.0). The framework makes extensive use of the \textsc{Eigen} linear-algebra library~\cite{ref:Eigen}, relying on fixed- and dynamic-size matrices and vectors for numerical efficiency and a uniform representation of kinematic quantities and covariance matrices throughout the package. Natively implemented for forward-geometry detectors, it can be applied directly to detectors with $4\pi$ coverage through appropriate state transformations.
The software is highly modular: the \ftransport, \fdoca, \fvertexer, and \ftreefitter components can be used either as an integrated reconstruction chain or independently in combination with external tools and algorithms. The package is currently used to reconstruct a wide variety of decay topologies at \lhcb and is being progressively incorporated into the experiment reconstruction and analysis software stack~\cite{LHCb:2023hlw}.

The paper is organised as follows. Section~\ref{sec:scenario} introduces the use case, including the main features of the \lhcb detector and its tracking system, as a representative example of the experimental scenario considered here.
Section~\ref{sec:fitter} discusses the key elements of the decay tree reconstruction algorithm, based on a simultaneous Kalman filter technique~\cite{Kalman:1960mft,Simon:2006} widely used in particle physics experiments since the 1990s~\cite{Billoir:1985nq,Fruhwirth:1987fm,Fruhwirth:1990zas, Hulsbergen:2005pu,Fruhwirth:2020zbo}. 
The algorithms used to provide optimal initialisation, a key requirement for efficient and reliable convergence, are described in Sec.~\ref{sec:seeding}. 
The performance of the framework is presented in Sec.~\ref{sec:performance} and compared with standard tools, using the benchmark decay channels $\Lb\to\jpsi\Lz$ and $\Bd\to\jpsi\KS$, with $\jpsi\to\mup\mun$, $\Lz\to\pr\pim$, and $\KS\to\pip\pim$, simulated and reconstructed within the \lhcb experiment. 
Detailed supplementary documentation is available online~\cite{ForwardTreeFitter2025:website}. It includes the master formulae of the Kalman-filter technique, explicit expressions for the decay chain constraints, the particle transport engines and forward- to/from $4\pi$-geometry state transformations, the simplified Kalman-based vertex fit used to initialise the simultaneous Kalman fit, a description of the code structure, the main configuration options, the \lhcb-specific settings used for the benchmark decay chains (representative of other decay chains), and a \textsc{ROOT}~\cite{ROOT_NIMA_1997} usage example.

\section{Use case}
\label{sec:scenario}

\begin{figure*}[!htb]
\centering
{\includegraphics[width=0.8\textwidth]{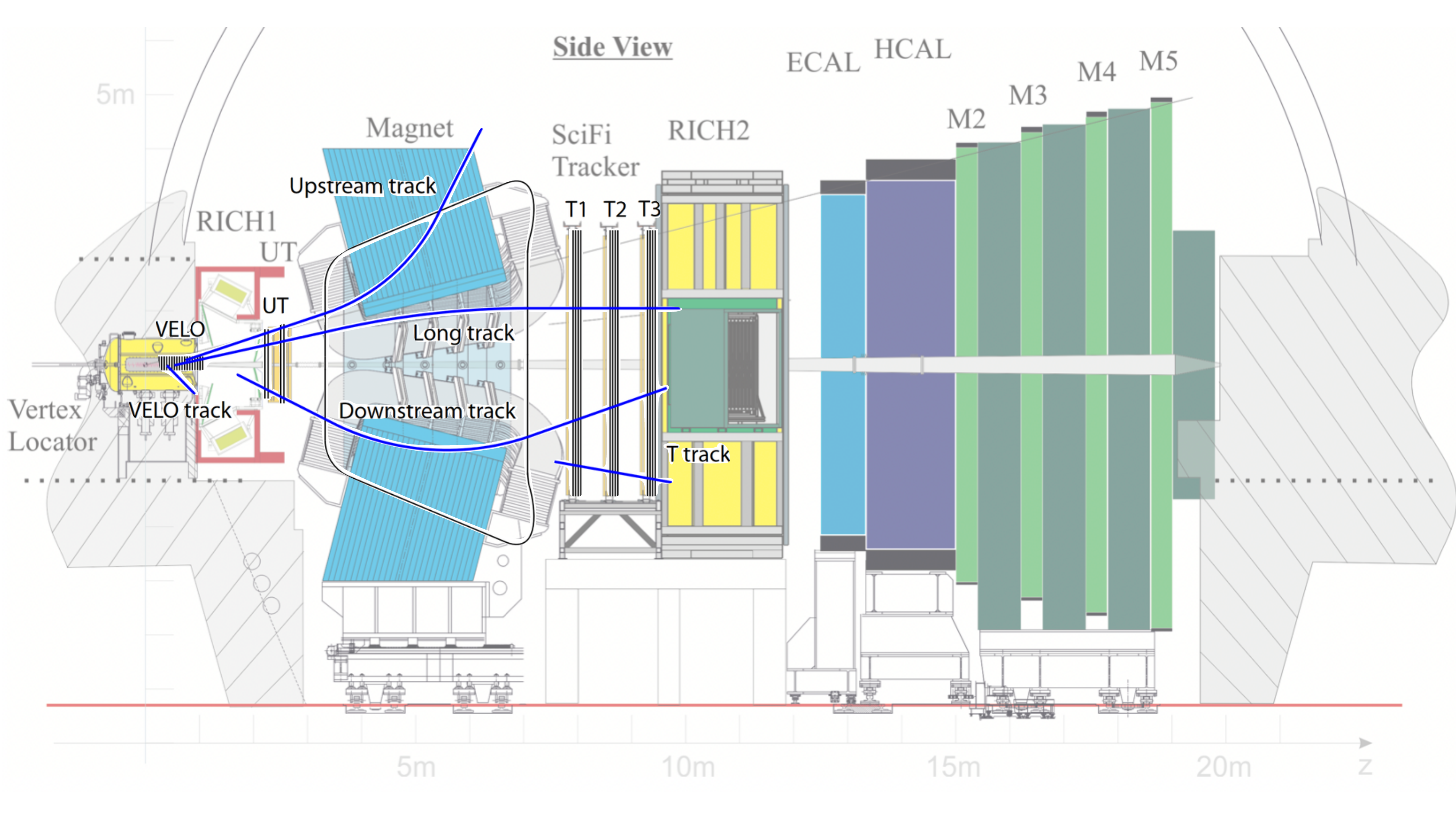}}
\caption{\lhcb detector layout, side view~\cite{LHCb:2023hlw}. The planes of the \lhcb tracking system, composed by the VELO, UT and SciFi Tracker, and the track-type definitions, 
are overlaid. For clarity, the overlaid tracks are shown in the $xz$ (bending) plane, \ie rotated by $90^\circ$ about the $z$ axis with respect to the detector layout, which is shown in the
$yz$ (non-bending) plane.
}
\label{fig:LHCb_sketch}
\end{figure*}

Figure~\ref{fig:LHCb_sketch} illustrates the \lhcb detector~\cite{LHCb:2023hlw}, highlighting its segmented tracking system and the resulting track categories. Track reconstruction
combines measurements from one or more of the Vertex Locator (VELO), surrounding the interaction region, the Upstream Tracker (UT), upstream of the dipole magnet, and the Scintillating Fibre (SciFi) Tracker, downstream of it. No tracking instrumentation is present in the extended decay volume between these detector systems.
Long tracks traverse the full tracking system, with hits in the VELO and SciFi and optionally in the UT, while Upstream tracks are reconstructed from VELO and UT hits and VELO tracks from the VELO alone. Downstream tracks combine UT and SciFi hits, whereas T tracks are reconstructed exclusively in the SciFi. For the latter, measurements are therefore available only at the SciFi stations,
with a separation of more than 5\,\m between the last UT layer and the first SciFi station. Decay vertices and kinematic properties within this large uninstrumented region must consequently be inferred by extrapolating the measured daughter trajectories upstream over distances of up to several metres, through the region where the magnetic field is strongest and most inhomogeneous. Downstream and T tracks are particularly relevant for long-lived particles decaying beyond the VELO, such as $\Lz$ and $\KS$ hadrons.

The progressively more downstream track categories extend the accessible decay region, from about 40\,\cm from the IP using Long tracks to 2.5\,m with Downstream tracks and about 7.5\,m with T tracks. For $\Lz$ hadrons from $\Lb\to\jpsi\Lz$ decays produced at the LHC, this corresponds to acceptances of about 11\%, 34\%, and 55\%, respectively, but at the cost of momentum resolutions of about 1\%, 2\%, and 20\%~\cite{LHCb:2022kwc}. In particular, the limited momentum resolution of T tracks results from the short lever arm and the relatively weak return magnetic field in the SciFi region, compared with the much longer trajectory through the magnetic field available for Long and Downstream tracks, as illustrated in Fig.~\ref{fig:LHCb_sketch}.

\begin{figure}[!htb]
\centering
{\includegraphics[width=0.50\textwidth]{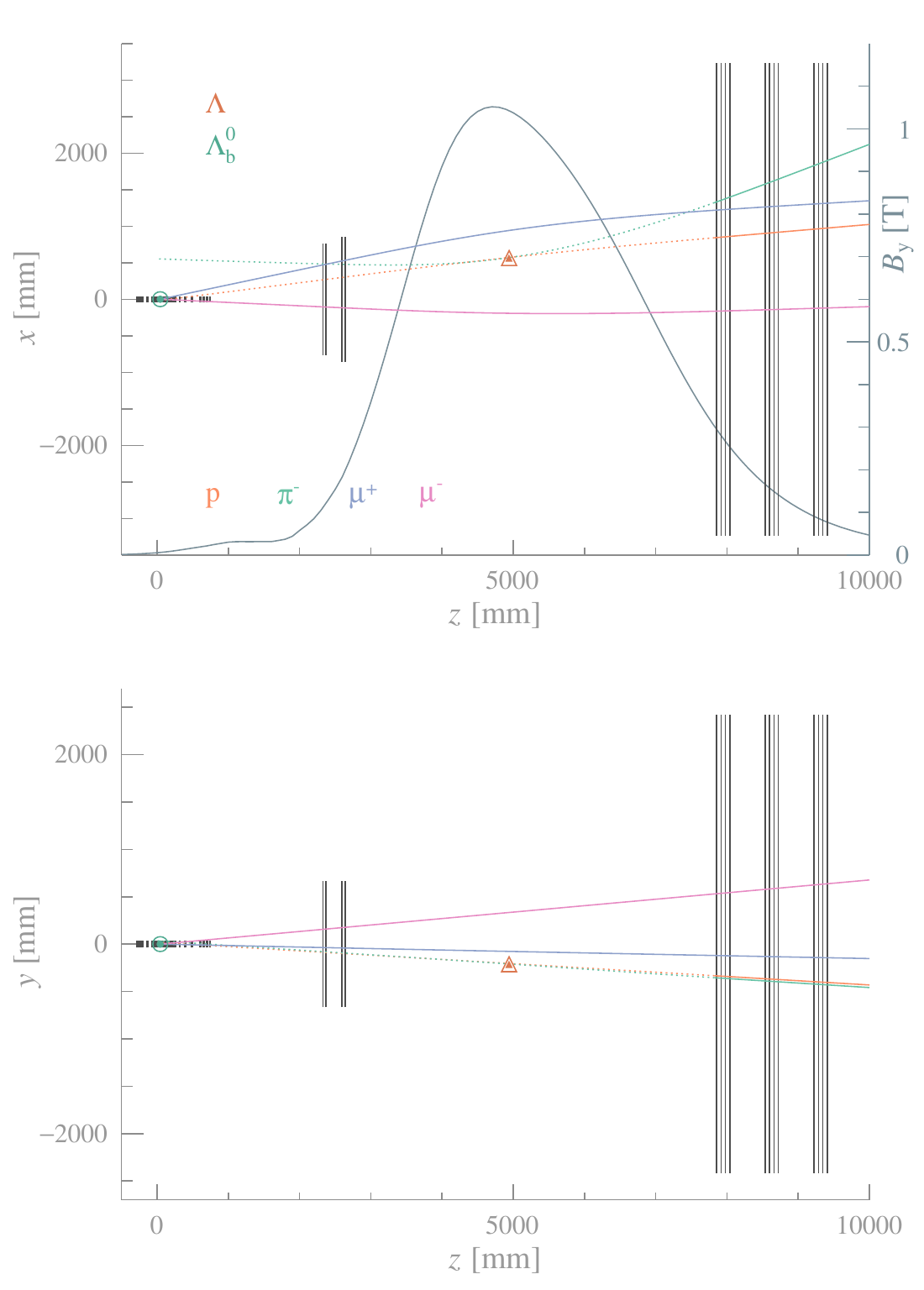}}
\caption{Event display of a simulated and reconstructed $\Lb\to\jpsi\Lz$, $\jpsi\to\mup\mun$, $\Lz\to\pr\pim$ decay chain in the (top) bending and (bottom) nonbending planes of the \lhcb detector. The trajectories of the final-state charged particles are shown as continuous lines downstream (from left to right) of their first hit measurement, while the dotted lines represent the trajectory extrapolations upstream, required for vertexing and kinematic fitting. The vertical lines represent sensitive planes (layers) of the \lhcb tracking system, VELO, UT and SciFi from left to right. The main magnetic-field component along the $y$ direction induced by the 4 T\,m dipole magnet is also overlaid in the bending plane. 
The open (filled) markers indicate the true (reconstructed) decay vertices of the \Lz (marron triangles) and \Lb (green circles) baryons.
}
\label{fig:Lb2JpsiL0_display}
\end{figure}

Figure~\ref{fig:Lb2JpsiL0_display} displays a simulated $\Lb\to\jpsi\Lz$, \mbox{$\jpsi\to\mup\mun$}, $\Lz\to\pr\pim$ decay chain reconstructed with the \lhcb detector, shown in the $zx$ (bending) and $zy$ (nonbending) planes, overlaid with the tracking layers and the main magnetic-field component along the $y$ direction induced by the 4 T\,m dipole magnet. The \Lb baryon originates at the primary vertex (PV) of the LHC proton-proton (\pr\pr) collision.
The \textit{detached} \Lz baryon originates at the \Lb decay vertex and traverses several metres inside the magnetic field before decaying into a proton and a pion in this example at approximately 5\,m from the IP, where the magnetic field is both maximal and inhomogeneous. The proton and pion are reconstructed as T tracks. 
Their trajectories must therefore be extrapolated upstream over several metres, from the first measurement plane located about 8\,m from the IP to the \Lz decay vertex, through an intense and strongly inhomogeneous magnetic field region, where the decay itself also takes place.
In contrast, the \jpsi meson can be treated as \textit{prompt}, with its decay vertex assumed to coincide with the \Lb decay position. Its decay muons are reconstructed as Long tracks with excellent momentum and position resolution. 
Particles are considered \textit{prompt} if their decay and production vertices are experimentally indistinguishable within detector resolution. For current detector technologies, this typically corresponds to a \mbox{$c\tau < 1\,\mum$}. 

This decay topology suggests stringent geometrical and kinematic constraints that can be exploited. In particular, the proton-pion and dimuon pairs must originate from common vertices, the \Lz momentum vector must align with its flight direction defined by the displacement vector between the \Lb and \Lz decay vertices, and the \Lb momentum vector must point back to the PV. In addition, the invariant masses of the \jpsi and \Lz candidates can be constrained to their known values~\cite{ParticleDataGroup:2024cfk}. Their optimal treatment within a simultaneous fit significantly improves the momentum and vertex resolutions.

\section{Simultaneous Kalman fit}
\label{sec:fitter}

The reconstruction strategy is based on a simultaneous least-squares fit of the full decay chain using the Kalman-filter formalism~\cite{Kalman:1960mft,Simon:2006,Billoir:1985nq,Fruhwirth:1987fm,Fruhwirth:1990zas,   Hulsbergen:2005pu,Fruhwirth:2020zbo}. In this approach, the minimisation of the global $\chi^2$ is carried out by adding the $\chi^2$ contributions from the individual constraints sequentially, rather than minimising the total $\chi^2$ in a single step~\cite{Frodesen:1979fy,Brandt:1976zc}. 
The fit combines reconstructed charged-particle trajectories, 
geometrical constraints, kinematic constraints, and optional invariant mass constraints into a common optimisation procedure that simultaneously determines the full set of decay chain parameters and their covariance matrix. 

\subsection{Scope and methodology}
\label{sec:scope_and_methodology}

In conventional decay tree fitting implementations, track parameters and their covariance matrices are transported from the measurement surface to the current estimate of the decay vertex at each iteration, where the measurement residuals are evaluated using local linear trajectory approximations around the vertex. This strategy is generally well suited for short extrapolation distances and weak magnetic fields. However, in the topologies targeted by the \ftreefitter\ algorithm, hereafter also referred to as \ftf, trajectories may need to be extrapolated over several metres through intense and strongly inhomogeneous magnetic field regions, while the decay vertices themselves may also lie inside these regions. Under these conditions, the repeated transport of track measurements and covariance matrices can become inaccurate or numerically unstable, and the local linear approximation of the particle trajectories around the decay vertex may no longer provide an adequate description, particularly when the momentum resolution is limited and non-Gaussian and when tight kinematic constraints, such as invariant mass constraints, are imposed.

To overcome these limitations, \ftf\ adopts a \emph{global track model} in which the track measurements remain fixed at their measurement surfaces, while the fitted track states are iteratively propagated through the magnetic field to these surfaces for the evaluation of the measurement residuals. This separates the track measurement from the \emph{trajectory model}, \ie the functional representation used to propagate the fitted state between the decay vertex and the measurement surface. 
The reference implementation employs a full numerical trajectory model with Runge--Kutta (RK) integration through the magnetic field map. In addition, the framework provides a range of \emph{local track models}, in which the track measurements are transported to the current estimate of the decay vertex, where the residuals are evaluated. Linear, parabolic and RK trajectory parameterisations are available, together with linear and interpolation-based transport engines, allowing the trade-off between computational performance and modelling accuracy to be adapted to the application.

The same principle applies to the treatment of exact constraints, including vertex, momentum conservation, and invariant mass constraints. Although, in the linear approximation, they can formally be interpreted as additional measurements, this description becomes increasingly inadequate for the topologies targeted by \ftf. Instead of linearising both the trajectory and constraint models simultaneously, \ftf\ preserves the full nonlinear trajectory model and re-linearises only the exact constraints around the updated solution at each iteration. This approach improves numerical stability while retaining the flexibility to incorporate arbitrary exact constraints.

The simultaneous Kalman fit proceeds iteratively until convergence of the state vector and covariance matrix is achieved. Dedicated seeding strategies are employed owing to the sensitivity of the convergence to initial conditions in the targeted topologies. Initial estimates of the decay vertices and parent momenta are obtained from pairwise distance- and point-of-closest-approach (DOCA/POCA) calculations combined with a simplified vertex Kalman fit. The corresponding covariance matrices are inflated and their correlation terms configured to stabilise convergence and prevent the initial seed from dominating the fit. The framework also supports flexible ordering and grouping of constraints to optimise numerical stability and convergence. 

\subsection{Constraints}
\label{sec:fitter_constraints}

The use case introduced in Sec.~\ref{sec:scenario} illustrates two types of constraints on the decay tree fit.
The first correspond to `measurement constraints' including the measured values from the four final-state particles (\pr, \pim, \mup, and \mun tracks), as well as the origin vertex of the head, \ie the \Lb, which must originate at the reconstructed PV position. A five-dimensional track state vector \mbox{$\bm s (z)^\meas \equiv (\bm r, \bm t, \psi)^\meas$} is the natural representation of the track measurement for a forward-geometry experiment such as \lhcb,
where the transverse position, \mbox{$\bm r^\meas \equiv (x,y)^\meas$}, slopes with respect to the $z$ axis, \mbox{$\bm t^\meas \equiv (t_x,t_y)^\meas$}, and the charge/momentum ratio, \mbox{$\psi^\meas \equiv (q/p)^\meas$},
with $q^\meas$ expressed in units of the positron charge,
are defined at a reference $z^\meas$ surface in the experiment. This reference position usually corresponds to the location along the $z$ axis of the detector plane containing the first measurement hit of the track or the point closer to the nominal beam axis~\cite{LHCb-2007-007}. In \lhcb, the former convention is adopted for Downstream and \mbox{T tracks}, whereas for Long tracks the choice is instead the latter. The superindex $\meas$ stands for `measurement', here and in the following. In this way, `track measurement' denotes the reconstructed state and its covariance at the reference surface. 
For the origin vertex constraint, \ie  `origin measurement',
\begin{eqnarray} 
  \label{eq:origin_constr}
\bm d & = & \bm o^\meas + \theta \bm p,
\end{eqnarray}
where \mbox{$\bm d=(x,y,z)$} and \mbox{$\bm o^\meas=(o_x,o_y,o_z)^\meas$} denote the decay vertex parameters and the origin vertex measurement, respectively, \mbox{$\bm p=(p_x,p_y,p_z)$} is the momentum vector of the decaying particle, $\theta = l/|{\bm p}|$ is a proper decay time parameter, with $l$ the decay length, and 
$p$ is the momentum magnitude.
With $l$ and $\bm p$ expressed in \mm and \mev, the decay time in seconds is directly $\theta m_0 /c$, where $m_0$ is the mass of the decaying particle expressed in \mev and $c$ the speed of light in $\mm\,{\rm s}^{-1}$. Note that~(\ref{eq:origin_constr}) assumes a linear trajectory for the decaying particle, which occurs either when the particle is electrically neutral or when the integrated magnetic field along the decay length is negligible. 
For \textit{prompt} decaying particles, it reduces to \mbox{$\bm d = \bm o^\meas$}. 
These constraints are of $\chi^2$-type, thus require the knowledge of the corresponding covariance matrices. Overall, there are $5 n_\trkbare = 20$ (23) measurement constraints without (with) an origin constraint, with $n_\trkbare=4$ the number of tracks.

The second type, referred to as `exact constraints', are applied to remove redundant degrees of freedom (`internal') and to constrain the decay kinematics (`external')~\footnote{Note that the meaning of external constraints here, as those exact constraints that are not internal, is different to Ref.~\cite{Hulsbergen:2005pu}, where they are equivalent to measurement constraints.}. The internal constraints include four-momentum conservation at the decay vertex, relating the parent particle and its decay products,
\begin{eqnarray} 
  \label{eq:4mom_constr}
(\bm p,E)_\beta & = & \sum_{i_\beta} (\bm p,E)_{i_\beta} + \sum_{\alpha_\beta} (\bm p,E)_{\alpha_\beta},
\end{eqnarray}
and the vertex constraint linking the parent particle to the production vertices of its children,
\begin{eqnarray} 
  \label{eq:decay_vertex_constr}
\bm d_{\alpha_\beta} & = & \bm d_\beta + \theta_{\alpha_\beta} \bm p_{\alpha_\beta}.
\end{eqnarray}
Here and in the following, $\beta$ refers to the parent particle, 
while $\alpha_\beta$ and $i_\beta$ denote the intermediate composite and final-state (track)  particles originating from the decay of particle $\beta$.
Analogously to the origin constraint, for a \textit{prompt} particle $\alpha_\beta$ this equation reduces to \mbox{$\bm d_{\alpha_\beta} =\bm d_\beta$}. Eliminating $\theta_{\alpha_\beta}$, the constraint~(\ref{eq:decay_vertex_constr}) can equivalently be rewritten as
\begin{eqnarray} 
  \label{eq:decay_vertex_constr_notheta}
\bm r_{\alpha_\beta} & = & \bm r_\beta + \bm t_{\alpha_\beta} (z_{\alpha_\beta} - z_\beta),
\end{eqnarray}
where $\bm r_{\alpha_\beta(\beta)}$ and $\bm t_{\alpha_\beta} = (p_x/p_z, p_y/p_z)_{\alpha_\beta}$ are the composite transverse positions and slopes.
Thus, there are $4 n_{\rm com} + 2 n_{\rm detached} + 3 n_{\rm prompt} = 17$ internal constraints, where $n_{\rm com} = 3$ is the total number of composite states (including the head), and $n_{\rm detached} = 1$ ($n_{\rm prompt} = 1$) the number of \textit{detached} (\textit{prompt}) intermediate states, \ie composites excluding the head.

The four-momentum and vertex constraints given by~(\ref{eq:4mom_constr}) and~(\ref{eq:decay_vertex_constr_notheta}) make the standard assumption that the three-momentum components of all child composites are identical at their production and decay points, and that their trajectories can be approximated as linear. These conditions strictly apply to \textit{prompt} or neutral child composites, and to charged \textit{detached} composites when the integrated magnetic field along the flight distance is negligible.
For charged long-lived composite particles such as $\Xi^-$, $\Sigma^\pm$, or $\Omega^-$ hyperons, these assumptions may become inaccurate even in weak magnetic fields. A parabolic `trajectory model' can be used for charged \textit{detached} child composites instead of the linear approximation, capturing the leading-order curvature with a closed-form expression that enters the constraints and projection matrices analytically. 

The external constraints include restricting the invariant mass of the composite state $\alpha$ to its known $m_\alpha$ value,
\begin{eqnarray} 
  \label{eq:mass_constr}
E_\alpha^2 & = & | \bm p_\alpha |^2 + m_\alpha^2.
\end{eqnarray}
This is usually applied to the intermediate states (unless they are strong resonances), \ie the \jpsi and \Lz (although could also be applied to the head, the \Lb), resulting in a total of 19 exact constraints.

The choice of parameters describing the decay tree is different for final-state and composite particles. For each track $i$ it uses the three-momentum, 
$\bm p_i$,
while for each composite it is a seven-dimensional vector containing the decay position and four-momentum, 
$(\bm r,z,\bm p,E)_\alpha$. 
For a composite decaying into tracks, its transverse $\bm r_\alpha$ position corresponds to the transverse position of all child tracks at a reference given by the $z$ vertex component of the decaying particle. The track slopes and charge/momentum ratio from the track parameters are directly given by $\bm t_i = (p_x/p_z,p_y/p_z)_i$ and $\psi_i=q_i^\meas/p_i$.
Note that the energy (or mass) of final-state particles is not treated as a free parameter, but is instead assigned according to the decay tree hypothesis.
Overall, we have $3\times 4+7\times3 =33$~(34) free parameters. With mass constraints applied to the \jpsi and \Lz particles, this results in \mbox{$20~(23) + 19 - 33~(34) = 6$~(8)} degrees of freedom.

\subsection{Particle transport}
\label{sec:fitter_transport}

To cover all scenarios, \ftransport includes RK and linear extrapolators, as well as a cubic Hermite interpolator, outlined below. 

The general-purpose RK engine is designed for fast, precise, and highly configurable track propagation through the inhomogeneous magnetic field, given a five-dimensional state vector and its covariance matrix along the particle trajectory. The transportation algorithm is based on the classic Runge--Kutta--Nyström method~\cite{Fruhwirth:2020zbo,Hairer1993,Fruhwirth:1990zas,Bugge:1980iv,Myrheim:1979ng} with adaptive step size. This technique is well suited to charged particle motion in magnetic fields, as it directly integrates second-order equations and naturally accounts for local field inhomogeneities.
The magnetic field is evaluated by trilinear interpolation from a three-dimensional vector field grid provided at initialisation.
Numerical differentiation for the evaluation of transverse field gradients make use of a second-order accurate finite-difference approximation within the same magnetic-field grid box.

The RK engine simplifies to a trivial linear transportation model when the track transport can be assumed to be linear. This is the case when the integrated magnetic field, and therefore the particle bending, are negligibly small. Optionally, both RK and linear extrapolators can account for multiple scattering in air.

The third engine uses cubic Hermite interpolation~\cite{BurdenFaires2010,CubicHermiteSplineWiki} between two states, or between pairs of a list of states ordered by increasing $z$. Extrapolation is performed using the same cubic representation. The interpolated covariance matrix incorporates the cross-covariance between the end-point states, and assumes a smooth linear variation of $\psi$. This engine is particularly relevant when multiple scattering (beyond air) and/or energy loss are significant. In \lhcb this occurs for Long tracks due to their interaction with the VELO RF foil, a thin, corrugated aluminium-based alloy shield that separates the detector secondary vacuum from the LHC beam vacuum~\cite{LHCb:2023hlw}.
An additional advantage of this approach is the potential speed-up of the decay tree fitting, since it avoids repeated RK propagation and the associated multiple accesses to the magnetic-field map.

\section{Seeding}
\label{sec:seeding}

The fit parameter state vector for the $k$-th constraint, $\bm v_k$, 
has to be determined using an iterative procedure since, in general, the constraints are not linear in $\bm v$ and the targeted topologies introduce sizeable non-linearities.
The sequence updating $\bm v_k$ and its covariance matrix $C_k$ for all constraints is repeated as follows. The initial fit parameters, $\bm v_0$, are set to $\bm v_{n_{\rm constr}}$, where $n_{\rm constr}$ is the total number of constraints from the previous iteration step. $C_0$ is set to its initial (seed) value at each iteration. The iterative procedure is repeated until the variation of the total $\chi^2$ between two consecutive iterations changes by an amount smaller than a given tolerance or the maximum number of iterations is reached. An additional `smoothing' iteration is performed to evaluate with full precision the $C_k$ covariance matrix once the iterative procedure is finished.

Suboptimal initialisation of the fit parameters can lead to significant convergence inefficiencies and degraded resolution. Therefore, the determination of the initial parameters $\bm{v}_0$ and their covariance matrix $C_0$ is, in general, a key ingredient in the Kalman filter technique, which becomes critical in the topologies targeted by \ftf.
The framework provides the tools and flexibility needed to apply the optimal initialisation procedure for the specific detector geometry and decay topology. To this end, two main tools are provided, as discussed in the following.

\vskip0.2cm
\noindent
\textit{Distance-of-closest approach}
\vskip0.2cm

It can be assumed that the vertex lies approximately where the separation between the reconstructed trajectories is minimal. An initial estimate of vertex positions and momenta in the $\bm{v}_0$ vector can therefore be obtained by identifying the points along the child trajectories
at which their Euclidean separation is minimised, \ie the distance-of-closest approach (DOCA). For more than two child particles, these two points can be evaluated from the pair of trajectories with the maximum DOCA. The point-of-closest approach (POCA) is then defined as the average position of the two points. The three-momenta and energies of all child particles can then be evaluated at the point of their trajectory that is closest to POCA.

When all trajectories are approximately linear, the evaluation of the maximum DOCA and associated POCA is straightforward and can be done analytically. However, for charged particles that must be extrapolated over long distances through regions of intense and inhomogeneous magnetic field, the procedure requires more refined optimisation techniques combined with the use of the RK extrapolator or the Hermite interpolator. 

Two different search algorithms are employed in the \fdoca algorithm: Golden Section in one dimension and Downhill Simplex in two dimensions~\cite{10.5555/506164}. In the former, the search is performed along a common $z$ coordinate for the two trajectories, while the latter allows for independent $z$ positions. In both cases, the algorithm proceeds iteratively by transporting the trajectory states from the previous to the current $z$ position(s), until the variation in the three-dimensional distance (which corresponds to the transverse distance in the Golden Section method) between two consecutive iterations falls below a given tolerance or the maximum number of iterations is reached.
In addition to the geometric DOCA, a transverse compatibility $\chi^2$ is evaluated from the Mahalanobis squared distance~\cite{Mahalanobis1936,McLachlan1999} between the two trajectory positions, using the propagated covariance matrices at the solution point. For the Golden Section algorithm, where the two trajectories share a common $z$ coordinate, an estimate of the uncertainty on the reconstructed POCA $z$ position is also provided from the local curvature of the Mahalanobis-distance profile around the minimum.
Composites are treated as tracks after transforming their vertex parameters $(\bm r, z, \bm p)\alpha$ to the standard $(\bm r, \bm t, \psi)\alpha$ track representation at $z_\alpha$.

\vskip0.2cm
\noindent
\textit{Vertex Kalman fit}
\vskip0.2cm

The computation of the DOCA does not make use of the track covariance
matrices, denoted as $V_k$. This has the advantage that the vector
$\bm v_0$ is robust against potential mismodelling of the reconstruction
uncertainties. Its main drawback is that it does not provide initial
estimates of the uncertainties and correlations of the vertex positions
and four-momenta needed to initialise $C_0$. To obtain a good
approximation to these expected covariance values, a Kalman fit
algorithm for fitting a single vertex at a time, \fvertexer, is
integrated into the framework.
The adopted strategy is then to set $C_0$ by scaling these expected
values by a configurable factor, chosen large enough that the initial
parameters carry negligible weight in the fit, but small enough that the
measurement errors $V_k$ are not numerically negligible.
In practice this factor is
typically in the range 100--10000, and the correlation terms can also be
configured to further improve convergence. 

\section{Performance}
\label{sec:performance}

The performance of the \forwardtreefitter algorithms is evaluated in terms of invariant mass resolution, vertex resolution along the $z$ axis, and reconstruction efficiency for decay chains with metre-scale displaced vertices. Discussion of the methodology and comparisons with the standard tools used in \lhcb are presented below.

\subsection{Procedure}
\label{sec:performance_procedure}

The performance of the algorithms integrated in the \forwardtreefitter framework has been benchmarked and compared with currently available \lhcb tools using simulated $\Lb\to\jpsi\Lz$ and $\Bd\to\jpsi\KS$ decays, with $\jpsi\to\mup\mun$, $\Lz\to\pr\pim$, and $\KS\to\pip\pim$.
The samples are generated using the standard \lhcb simulation framework~\cite{Sjostrand:2007gs,Belyaev:2011zza,Lange:2001uf,Golonka:2005pn,
Agostinelli:2002hh,Allison:2006ve,Clemencic:2011zza}\footnote{We specifically use \pythia~8.244, \evtgen~2.2.3 and \photos~3.56 to generate the $pp$ collisions, the decays of unstable particles and the final-state radiation, respectively. The detector response employs the \geant~10.6.2 toolkit.}, corresponding to the 2024 data-taking conditions~\cite{ref:MCsamples}, with an average of $4.4$ visible $pp$ interactions per bunch crossing, and are reconstructed using the standard \lhcb workflow~\cite{LHCb:2023hlw}.
These decay chains are chosen because they contain two representative long-lived hadrons in \lhcb and provide complementary topologies in which the parent particles have different masses, lifetimes, and $Q$-values~\cite{ParticleDataGroup:2024cfk}. The two channels also differ in their daughter kinematics: the \Lz decays to particles of unequal mass ($p\pi^-$), whose tracks therefore have different curvatures, whereas the \KS decays to a pair of equal mass ($\pi^+\pi^-$) with similar curvatures. The differing $Q$-values lead, in addition, to different opening angles. Other \bquark- and \cquark-hadron decay chains involving \Lz and \KS decays that have been studied show results consistent with those reported below.

The $\jpsi$ candidates are formed by combining opposite-charge Long tracks identified as muons, while \Lz and \KS candidates are reconstructed from pairs of two T, Downstream, or Long tracks with opposite charge (TT, DD, and LL pairs). The \Lz and \KS candidates are then combined with the $\jpsi$ to form \Lb and \Bd candidates. Fake candidates arise either from random combinations of tracks associated with genuine charged-particle trajectories, or from fake tracks originating from random hit combinations or also, in the case of Long and Downstream tracks, from mismatched track segments upstream and downstream of the magnet. Signal candidates are defined as those in which all tracks and composites are matched to the corresponding true particles, while background candidates are those for which both the \Lb/\Bd and \Lz/\KS combinations are fake. A very loose selection is applied to reduce the combinatorial background prior to running the \forwardtreefitter reconstruction sequence.
After this selection, the sample contains approximately twice as many \Lb and \Bd signal as background candidates.

For the benchmark \Lb (\Bd) decay tree, two reconstruction sequences are
executed, each consisting of a series of simultaneous Kalman fits run
sequentially and seeded from the results of previous stages.
In SEQ1, one \ftf configuration is applied to the \Lz (\KS) hadron
(inclusive mode), and four configurations are applied to the full decay
chain (exclusive mode): without mass constraints; with a mass constraint
on the \jpsi only; with mass constraints on both the \jpsi and \Lz
(\KS); and with the same constraints but exchanging the \KS (\Lz)
particle hypotheses to account for proton-pion misidentification. In
SEQ2, the complete \ftf sequence is cloned with the PV constraint enabled
and executed after SEQ1.

The seeding proceeds in stages. At the lowest level, \fdoca is used to determine the initial vertex positions and momenta for single vertex fits by \fvertexer. These vertex fits in
turn seed the simultaneous fits: the seed parameters of the exclusive \ftf
without mass constraints and of the inclusive \ftf, both without the PV
constraint, are taken from \fvertexer applied to the full decay chain and
to the \Lz (\KS) only, respectively. All other exclusive and inclusive
simultaneous fits are then initialised from the corresponding results without mass or PV constraints, with position and momentum covariance matrix
elements increased by a factor of 100, and vanishing off-diagonal terms,
except when a mass constraint on the \Lz (\KS) is applied. Multiple
scattering in air is disabled, as its effect is found to be negligible
(0.33\%$X_0$ per metre).

The simultaneous decay chain fit results are computed separately for two regions of the decay volume, as the mass resolution depends strongly on the \Lz (\KS) decay position along the $z$ axis, owing to the increasing lever arm of the daughter tracks at smaller $z$. The regions considered are \mbox{$3-6\,\m$} and \mbox{$6-8\,\m$}. This separation also reflects the fact that the \lhcb magnetic field reaches its maximum around 5\,\m.

\begin{figure}[!htb]
\centering
\begin{align}
{\includegraphics[width=0.245\textwidth]{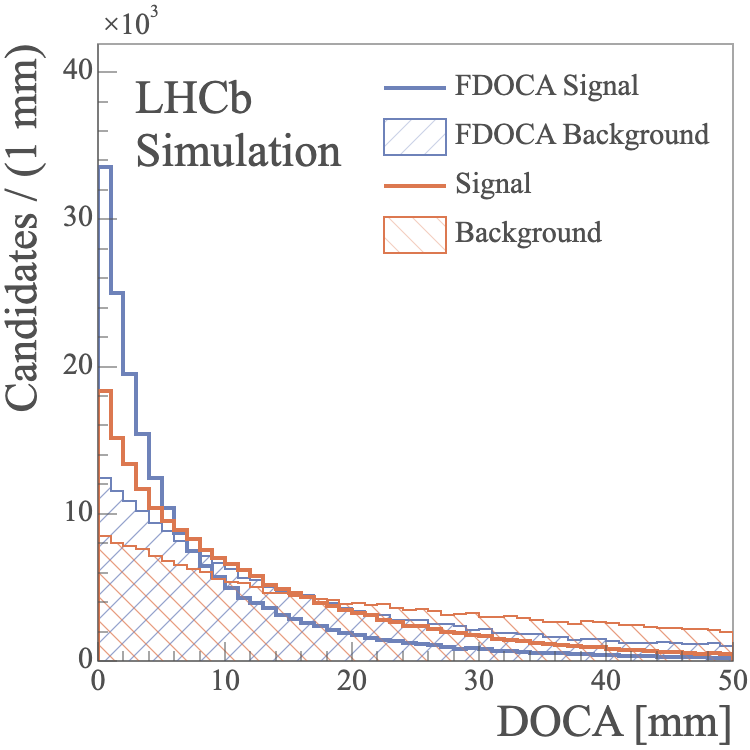}} & 
{\includegraphics[width=0.245\textwidth]{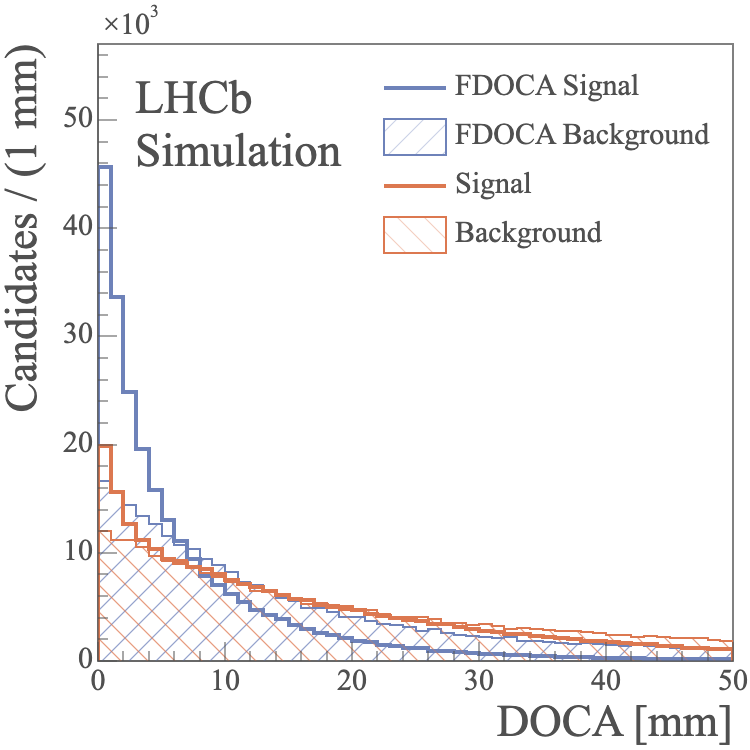}} \nonumber
\end{align} 
\caption{Distributions of the DOCA for final-state TT pairs from signal \Lz (left) and (right) \KS decays, reconstructed using \fdoca and the standard \lhcb tool. Background-pair distributions are overlaid and normalised to the signal.
}
\label{fig:DOCAs}
\end{figure}

\begin{figure}[!htb]
\centering
\begin{align}
{\includegraphics[width=0.245\textwidth]{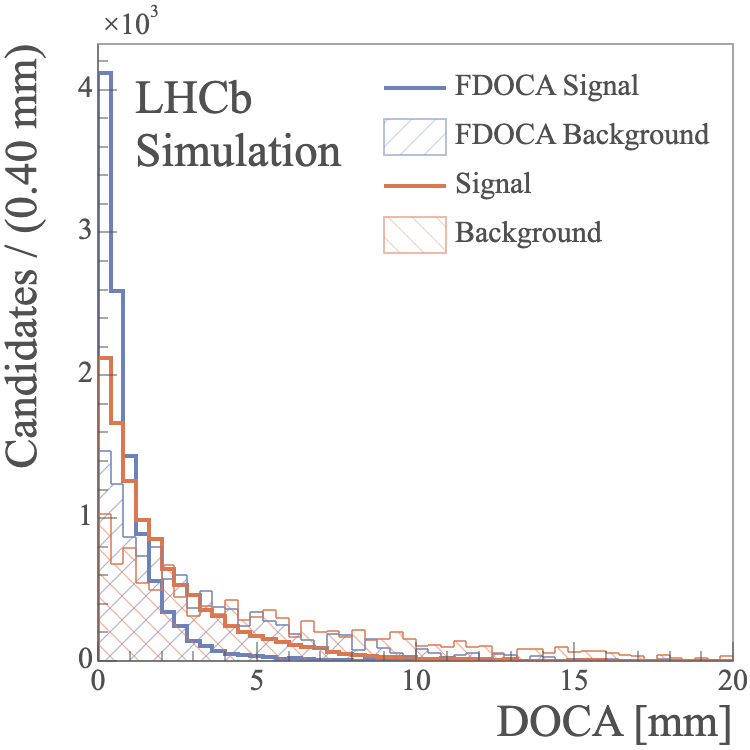}} &
{\includegraphics[width=0.245\textwidth]{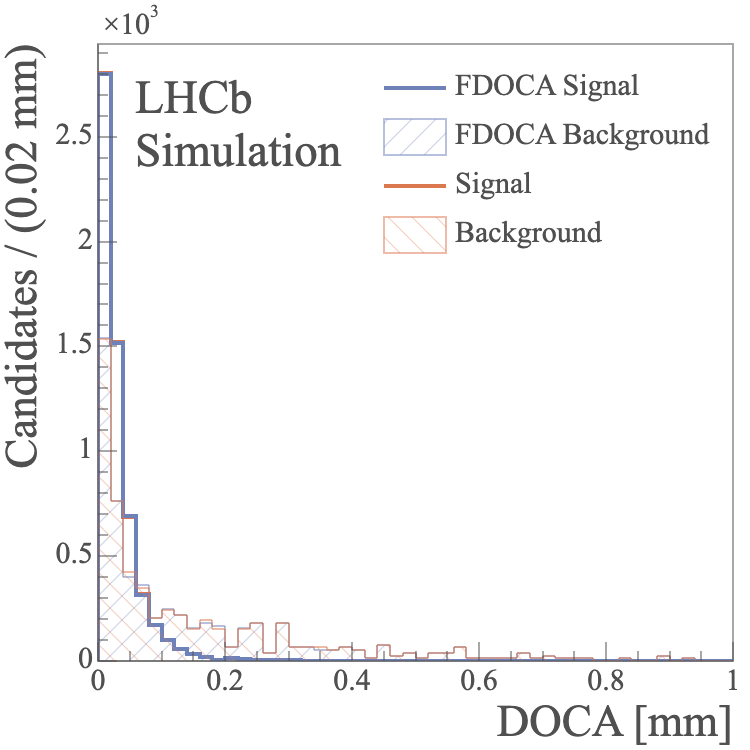}} \nonumber
\end{align} 
\caption{Distributions of the DOCA for final-state (left) DD and (right) LL pairs from signal \Lz decays, reconstructed using \fdoca and the standard \lhcb tool. Background-pair distributions are overlaid and normalised to the signal.
}
\label{fig:DOCAs_DD_LL}
\end{figure}

\subsection{Results}
\label{sec:performance_results}

\begin{table}[htbp]
\centering
\caption{Comparison of the invariant mass and $z$-vertex resolutions obtained with the standard \decaytreefitter\ (\dtf) and \ftf\ for the benchmark \Lb and \Bd decay chains, quoted in units of\,\mev and\,\mm, respectively. The \Lz and \KS candidates are reconstructed from TT track pairs in the two regions of the \lhcb decay volume, with mass constraints imposed on the \jpsi, \Lz and \KS candidates, and the PV constraint applied to the parent \Lb add \Bd hadrons. The last column gives the resolution improvement factor, defined as $\sigma_\dtf/\sigma_\ftf$.
}
\label{tab:resolutionsFTFvsDTF}
\begin{tabular}{lcccc}
\toprule
Observable & Range\,[m] & \dtf & \ftf & Improvement \\
\midrule
\multirow{2}{*}{\Lb\ mass}
    & $3-6$ & 150 & 12 & $12.5$ \\
    & $6-8$ & 100 & 43 & $2.3$ \\
\midrule
\multirow{2}{*}{\Bd\ mass}
    & $3-6$ & 95 & 16 & $5.9$ \\
    & $6-8$ & 53 & 40 & $1.3$ \\
\midrule
\multirow{2}{*}{\Lz\ $z$-vertex}
    & $3-6$ & 570 & 150 & $3.8$ \\
    & $6-8$ & 230 & 98 & $2.3$ \\
\midrule
\multirow{2}{*}{\KS\ $z$-vertex}
    & $3-6$ & 290 & 39 & $7.4$ \\
    & $6-8$ & 55 & 32 & $1.7$ \\
\bottomrule
\end{tabular}
\end{table}

When the mass constraints are imposed, \ftf substantially outperforms the current \decaytreefitter (\dtf)~\cite{Hulsbergen:2005pu} implementation in the LHCb software for candidates reconstructed from TT pairs,
as summarised in Table~\ref{tab:resolutionsFTFvsDTF}.
Here and in the following the resolution is defined as the central 68.3\% confidence-level (CL) interval of the residual distribution, defined as the difference between reconstructed and true values.
The reconstruction efficiency is improved
correspondingly: with the mass constraints applied, \ftf maintains an
approximately flat efficiency of $\sim\!96\%$ across the full
$3-8$\,m range, whereas \dtf loses efficiency sharply for decays
between $3.5-7$\,m, falling to about $64\%$ in the $4.5-6$\,m region
for both the \Lb and \Bd channels. Without the mass constraints, the
two algorithms yield broadly comparable resolutions and efficiencies. The improvement is driven by the \emph{global track model}, as well as the improved seeding and constraint ordering. These results are decomposed below.

The performance of \fdoca improves by $30\%$ for \Lz and $50\%$ for \KS decays from TT pairs  with respect to the standard \lhcb DOCA tool, as quantified  by the significance figure-of-merit ${\cal S} = N_\mathrm{sig}/\sqrt{N_\mathrm{sig}+N_\mathrm{bkg}}$, where $N_\mathrm{sig}$ and $N_\mathrm{bkg}$ denote the signal and background yields in the DOCA range up to 5\,\mm. Figure~\ref{fig:DOCAs} shows the corresponding DOCA distributions for signal and background reconstructed with \fdoca and with the standard tool, which assumes linear track trajectories. The signal distributions obtained with \fdoca are more sharply peaked near zero and exhibit improved separation from the background. 
A similar behaviour is observed for DD pairs. For \Lz candidates reconstructed from DD pairs, a gain in significance of about $30\%$ is obtained in the DOCA range up to 1\,\mm, while for LL pairs the two algorithms yield equivalent performance, as illustrated in Fig.~\ref{fig:DOCAs_DD_LL}.
The DOCA distributions for $J/\psi$ candidates are identical for \fdoca and for the standard tool.

\begin{figure}[!htb]
\centering
\begin{align}
{\includegraphics[width=0.245\textwidth]{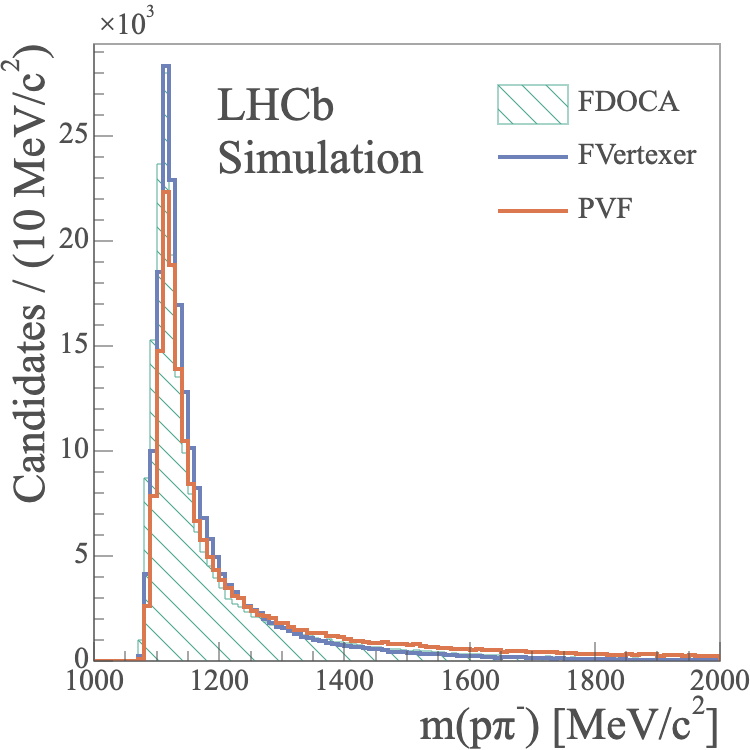}} & 
{\includegraphics[width=0.245\textwidth]{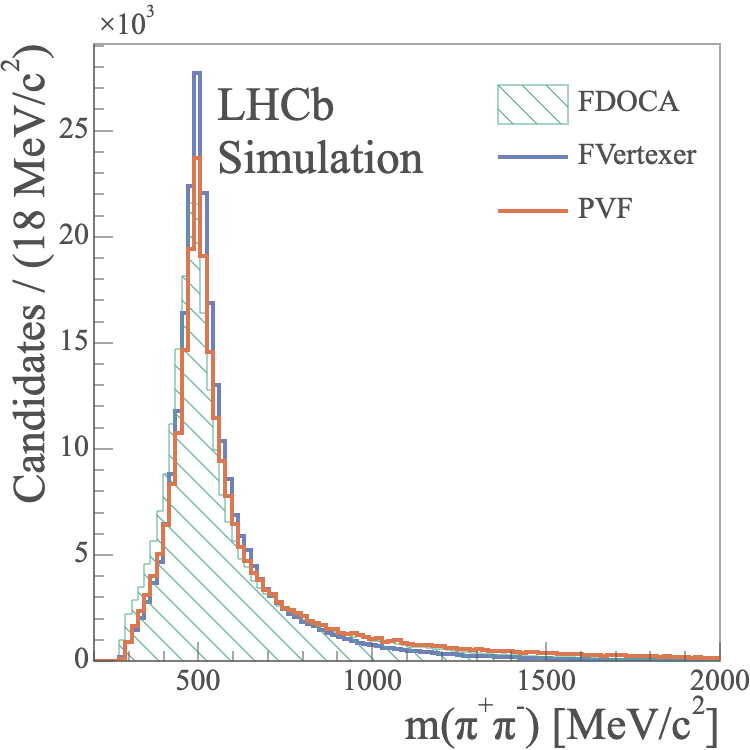}} \nonumber
\end{align}
\caption{Reconstructed invariant mass distributions of (left) \Lz and (right) \KS candidates from TT pairs, obtained using \fdoca, \fvertexer, and the standard \pvf.
}
\label{fig:Vertexer_LLPs}
\end{figure}

\begin{figure}[!htb]
\centering
\begin{align}
{\includegraphics[width=0.245\textwidth]{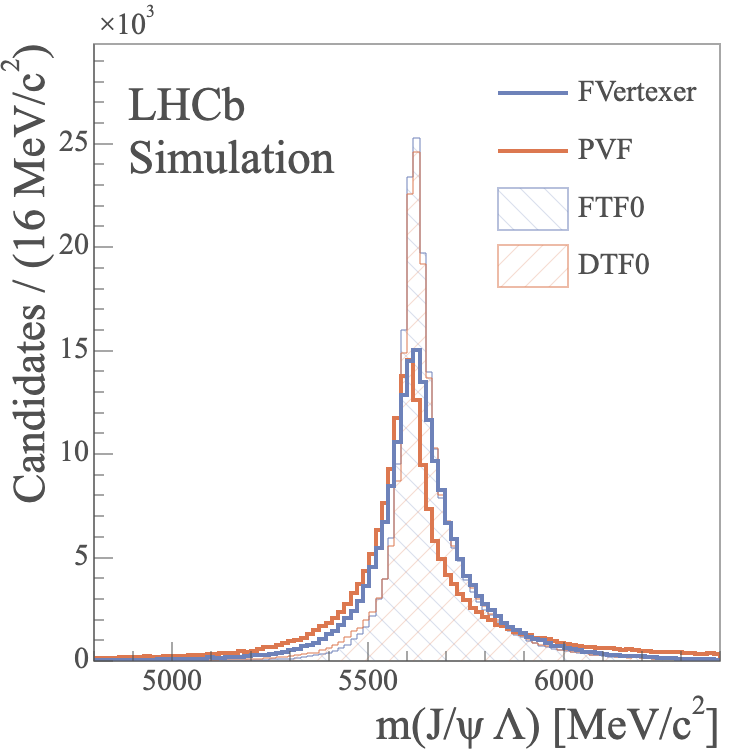}} &
{\includegraphics[width=0.245\textwidth]{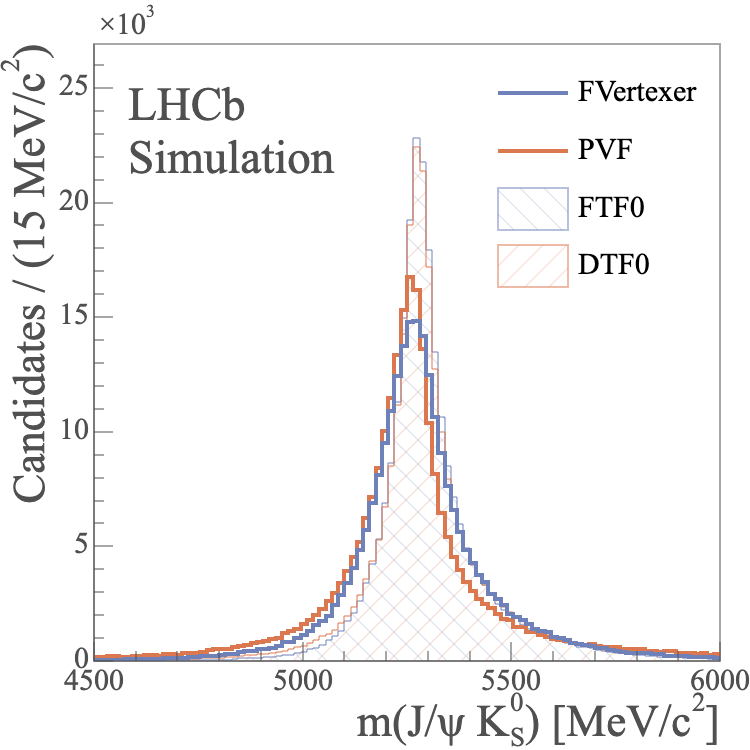}} \nonumber
\end{align}
\caption{Reconstructed invariant mass distributions of (left) \Lb and (right) \Bd candidates from TT pairs obtained with \fvertexer and \ftf with the PV constraint and no mass constraints (label 0),
compared with the standard \pvf and \dtf tools.
}
\label{fig:Vertexer_Bs}
\end{figure}

\begin{figure}[!htb]
\centering
\begin{align}
{\includegraphics[width=0.245\textwidth]{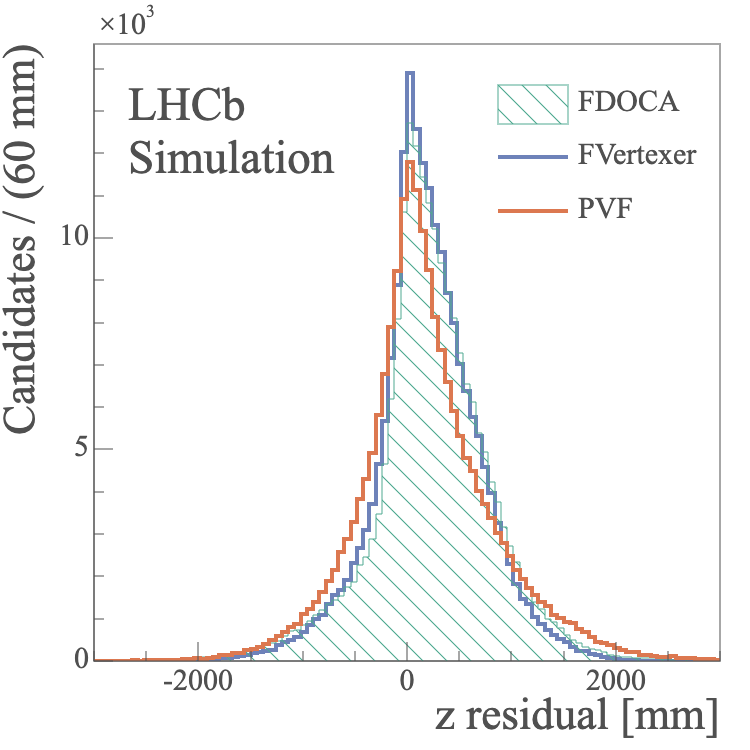}} &
{\includegraphics[width=0.245\textwidth]{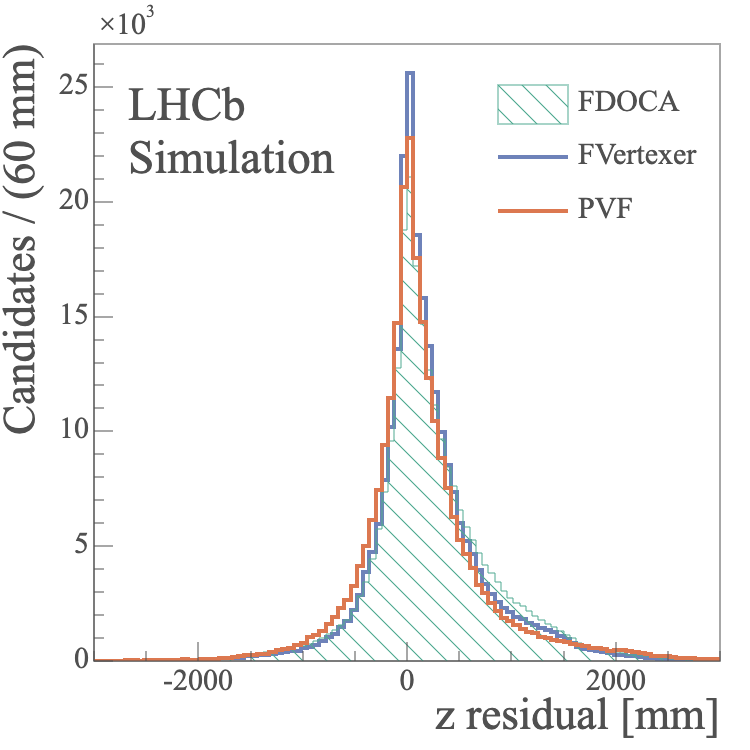}} \nonumber
\end{align}
\caption{Reconstructed $z$-vertex residual of (left) \Lz and (right) \KS candidates from TT pairs obtained using \fdoca, \fvertexer, and the standard \pvf.
}
\label{fig:zVertex}
\end{figure}

\begin{table}[htbp]
\centering
\caption{Comparison of the invariant mass and $z$-vertex resolutions obtained with the standard \pvf\ and \fvertexer\ for the benchmark \Lb\ and \Bd\ decay chains, quoted in units of\,\mev and\,\mm, respectively. The \Lz and \KS candidates are reconstructed from TT track pairs over the full \lhcb decay volume ($z>3$\,m). The last column gives the resolution improvement factor, defined as $\sigma_\dtf/\sigma_\ftf$.
}
\label{tab:resolutionsFVTXvsPVF}
\begin{tabular}{lccc}
\toprule
Observable & \pvf & \fvertexer & Improvement \\
\midrule
\Lz\ mass
    & 190 & 80 & 2.4 \\
\addlinespace
\KS\ mass
    & 200 & 130 & 1.5 \\
\addlinespace
\Lb\ mass
    & 170 & 120 & 1.4 \\
\addlinespace
\Bd\ mass
    & 140 & 130 & 1.1 \\
\addlinespace
\Lz\ $z$-vertex
    & 570 & 440 & 1.3 \\
\addlinespace
\KS\ $z$-vertex
    & 430 & 400 & 1.1 \\
\bottomrule
\end{tabular}
\end{table}

Using \fvertexer, the reconstruction performance improves with respect to the standard \particlevertexfitter\ (\pvf), as summarised in Table~\ref{tab:resolutionsFVTXvsPVF}. The largest gains are obtained for the reconstruction of the long-lived \Lz\ and \KS\ candidates, while more modest improvements are observed for the parent \Lb\ and \Bd\ invariant masses. The \pvf\ algorithm is optimised for real-time HLT2 reconstruction in \lhcb~\cite{Hulsbergen2018VertexToolsNote,LHCb:2023hlw}. The improved performance of \fvertexer\ is mainly attributable to the optimised vertex seed provided by \fdoca\ together with the iterative Runge--Kutta-based track transport, replacing the single RK-based initial track-state estimation currently implemented in \pvf\ for TT-track candidates.

The reconstructed \Lz\ (\KS) and \Lb\ (\Bd) invariant mass distributions obtained with \pvf\ and \fvertexer\ are compared in Figs.~\ref{fig:Vertexer_LLPs} and~\ref{fig:Vertexer_Bs}, respectively. For the long-lived \Lz\ (\KS) candidates, the masses reconstructed with \fdoca\ are also shown. Figure~\ref{fig:zVertex} compares the residual distributions of the \Lz\ and \KS\ $z$-vertex positions for the three algorithms. The reconstructed \jpsi\ invariant mass distributions are essentially identical for all three algorithms, and similarly good agreement is observed in the \Lz\ (\KS) and \Lb\ (\Bd) invariant mass and $z$-vertex residual distributions when the long-lived hadrons are reconstructed from DD or LL track pairs.


\begin{figure}[!htb]
\centering
\begin{align}
{\includegraphics[width=0.245\textwidth]{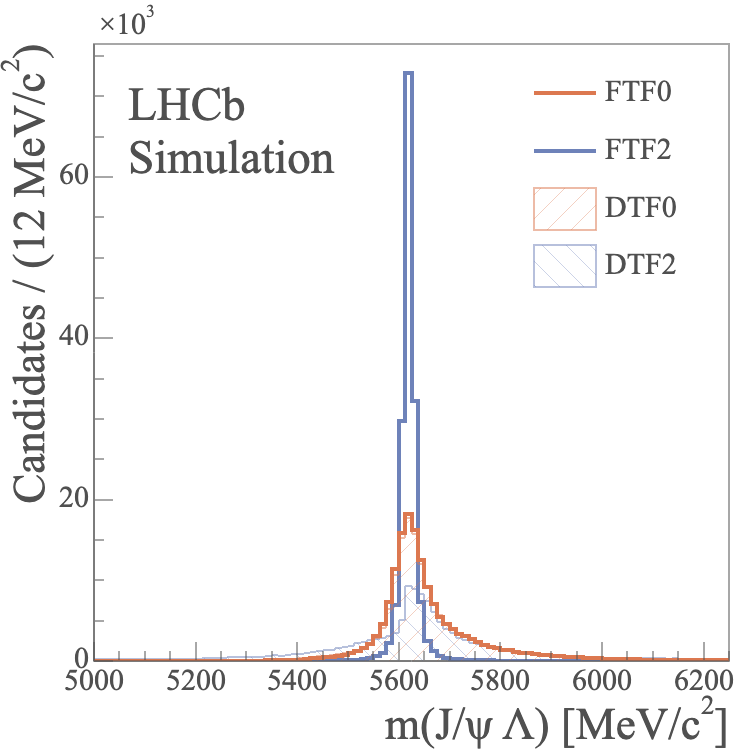}} &
{\includegraphics[width=0.245\textwidth]{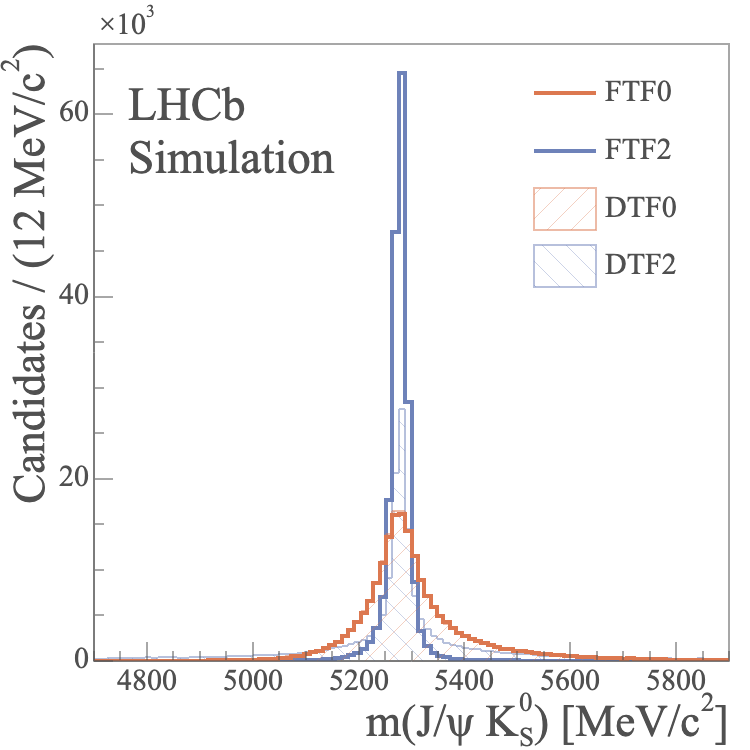}} \nonumber \\
{\includegraphics[width=0.245\textwidth]{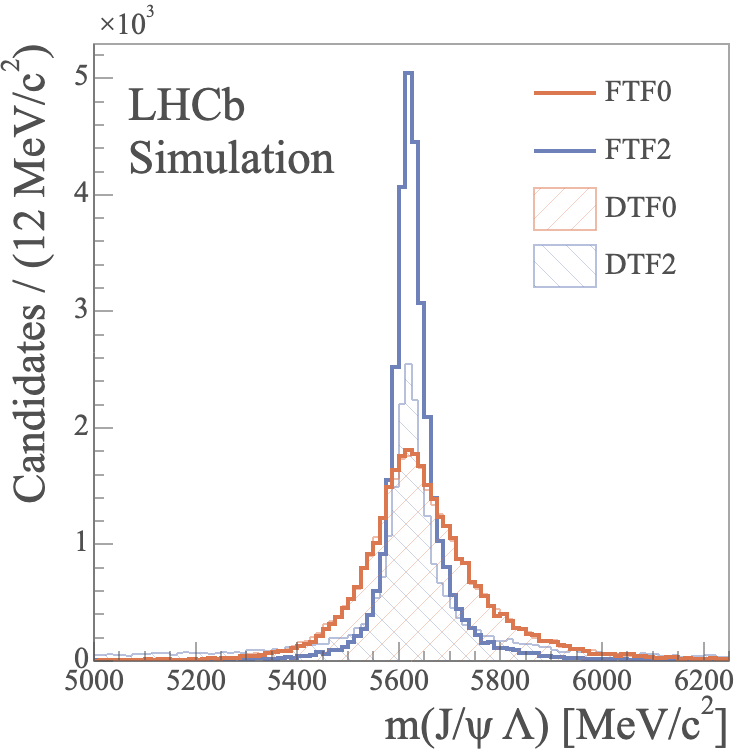}} &
{\includegraphics[width=0.245\textwidth]{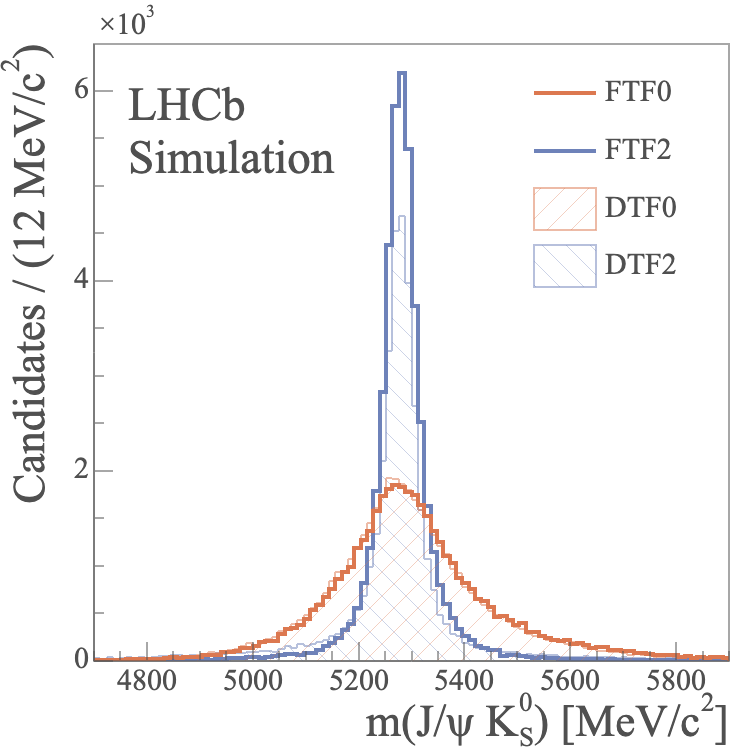}} \nonumber
\end{align}
\caption{(Left) \Lb and (right) \Bd invariant mass distributions reconstructed with \ftf and \dtf, using the PV constraint and either without (label 0) or with (label 2) the mass constraints applied to the \jpsi and long-lived hadrons from TT pairs. The results are shown separately for two regions of the decay volume: (top) \mbox{$3-6\,\m$} and (bottom) \mbox{$6-8\,\m$} along the $z$ axis.
}
\label{fig:FTF_mass}
\end{figure}

\begin{figure}[!htb]
\centering
\begin{align}
{\includegraphics[width=0.245\textwidth]{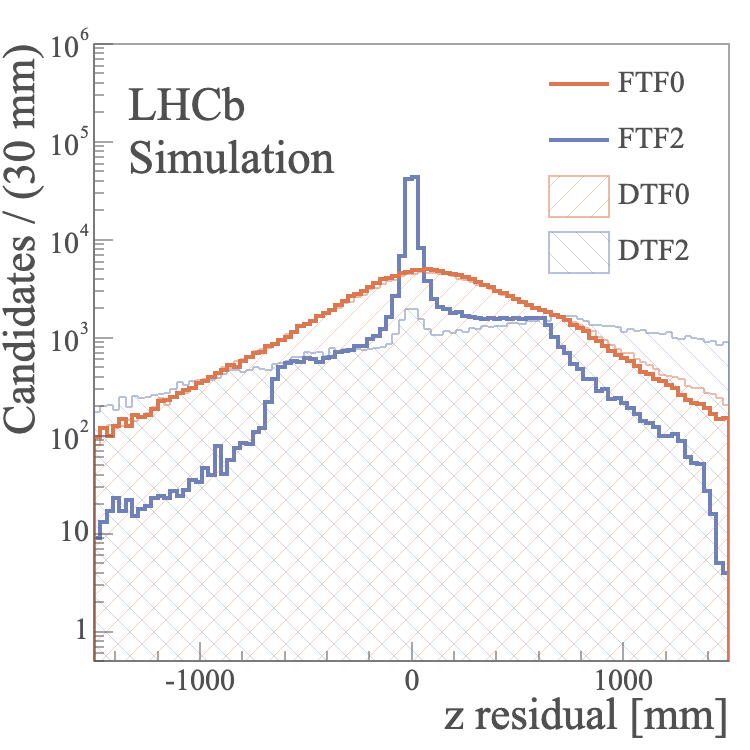}} &
{\includegraphics[width=0.245\textwidth]{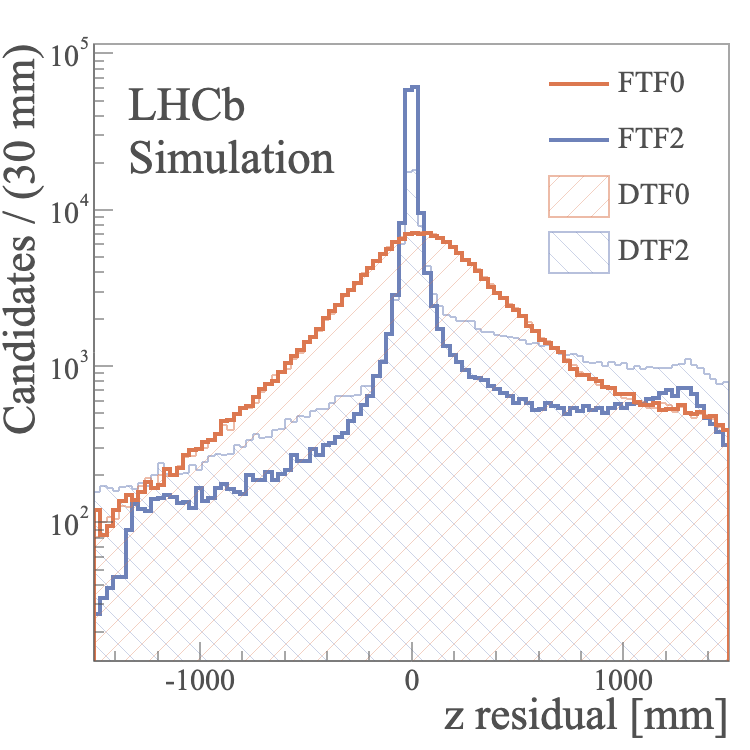}} \nonumber \\
{\includegraphics[width=0.245\textwidth]{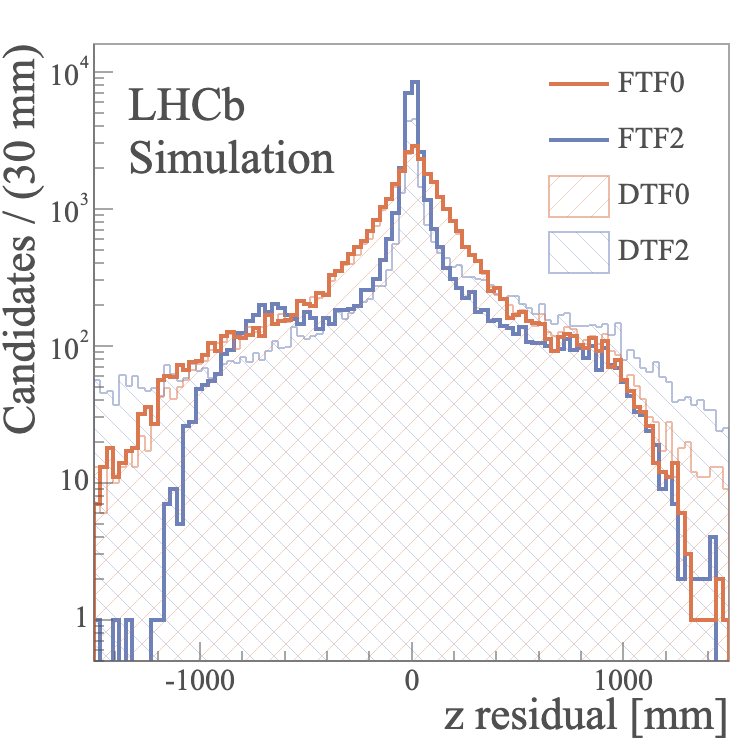}} &
{\includegraphics[width=0.245\textwidth]{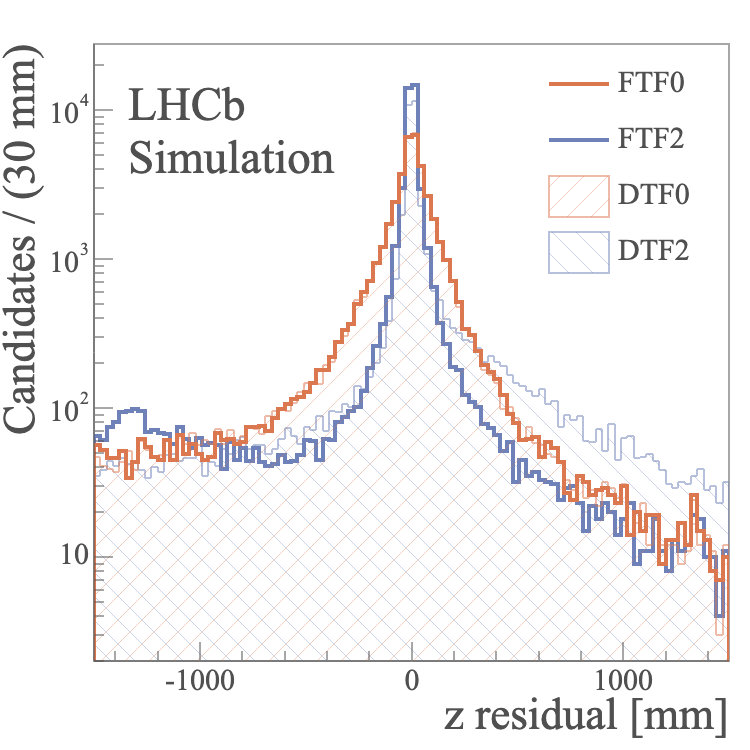}} \nonumber 
\end{align}
\caption{(Left) \Lz and (right) \KS $z$-vertex resolution obtained with \ftf and \dtf, using the PV constraint and either without (label 0) or with (label 2) the mass constraints applied to the \jpsi and long-lived hadrons from TT pairs. The results are shown separately for two regions of the decay volume: (top) \mbox{$3-6\,\m$} and (bottom) \mbox{$6-8\,\m$} along the $z$ axis.
}
\label{fig:FTF_zresol}
\end{figure}

The \Lb (\Bd) invariant mass distributions reconstructed with \ftf, using the PV constraint and either without or with mass constraints applied to the \jpsi and \Lz (\KS) hadrons, are shown in Fig.~\ref{fig:FTF_mass}, together with the corresponding distributions obtained using \dtf. 
The corresponding \ftf distributions using the PV constraint alone, integrated over the full decay range, are also shown and compared to the vertexing algorithms in Fig.~\ref{fig:Vertexer_Bs}.

Figure~\ref{fig:FTF_zresol} shows the corresponding $z$-vertex resolution of the long-lived \Lz (\KS) hadron across the decay volume. 
The residual distributions exhibit long and asymmetric non-Gaussian
tails, along with more localised broad structures at large positive
and negative values, both depending strongly on the decay position.
These large residuals are not accompanied by a corresponding increase
in the estimated vertex uncertainty, resulting in large pull values. 
The fraction of these large-residual events is significantly reduced by \ftf as compared to \dtf,
although it remains non-negligible, particularly for \Lz decays occurring closer to the IP region.
Both features originate from an intrinsic information limitation in
nearly collinear low-$Q$ decays. The poorly resolved angular opening
in the non-bending plane weakens the longitudinal vertex determination:
its nonlinear propagation gives rise to the asymmetric non-Gaussian
tails, while the same loss of information enhances the ambiguity
between possible geometrical crossings in the bending plane, producing
the broad structures.
A discussion of their dependence on the decay kinematics,
track resolution and decay position is given in Appendix~\ref{sec:performance_lowQ}.

Both the \Lb\ (\Bd) invariant mass and \Lz\ (\KS) $z$-vertex resolutions are substantially improved with \ftf\ with respect to \dtf\ when mass constraints are imposed, as summarised in Table~\ref{tab:resolutionsFTFvsDTF}. The improvement is driven primarily by the reconstruction of the long-lived hadron, particularly in the $3-6$\,\m region, where the extrapolation distances are largest and the magnetic field is strongest. The estimate of the $z$-vertex resolution uses events with residuals within a $\pm1000$\,\mm window.
Moreover, the fraction of events featuring large residual is significantly reduced, although it remains non-negligible, particularly for \Lz\ decays occurring closer to the IP region.
Without the mass constraints, the resolutions are broadly comparable. These results highlight the importance of the \emph{global track model}, combined with optimal seeding, when fitting tightly constrained decay trees with particles decaying in regions of intense and inhomogeneous magnetic field, in contrast to the standard approach based on linearised track models. The use of the \emph{parabolic} or \emph{RK local track models} already leads to substantial improvements over the \textit{linear local track model}, although the best performance is achieved with the nominal \emph{global track model}.

\begin{figure}[!htb]
\centering
\begin{align}
{\includegraphics[width=0.245\textwidth]{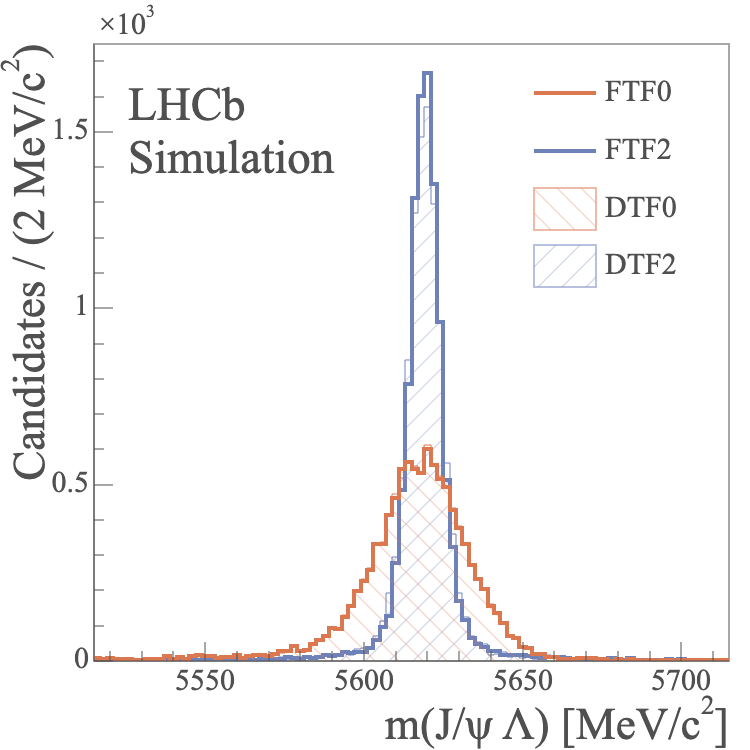}} &
{\includegraphics[width=0.245\textwidth]{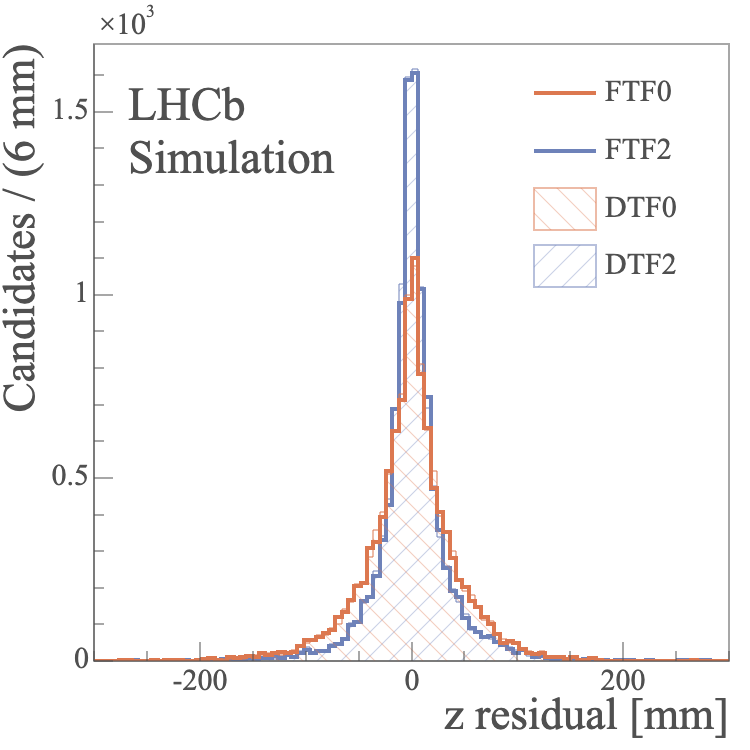}} \nonumber
\end{align}
\caption{(Left) \Lb invariant mass distribution and (right) \Lz $z$-vertex resolution for candidates reconstructed from DD pairs, obtained with \ftf and \dtf, using the PV constraint and either without (label 0) or with (label 2) the mass constraints applied to the \jpsi and \Lz hadrons.
}
\label{fig:FTF_mass_zresol_DD}
\end{figure}

The \Lb (\Bd) invariant mass and $z$-vertex residuals for \Lz (\KS) candidates reconstructed from DD and LL pairs with \ftf and \dtf are similar, as illustrated in Fig.~\ref{fig:FTF_mass_zresol_DD} for the case of \Lb decays with DD pairs.

\begin{figure}[!htb]
\centering
\begin{align}
{\includegraphics[width=0.245\textwidth]{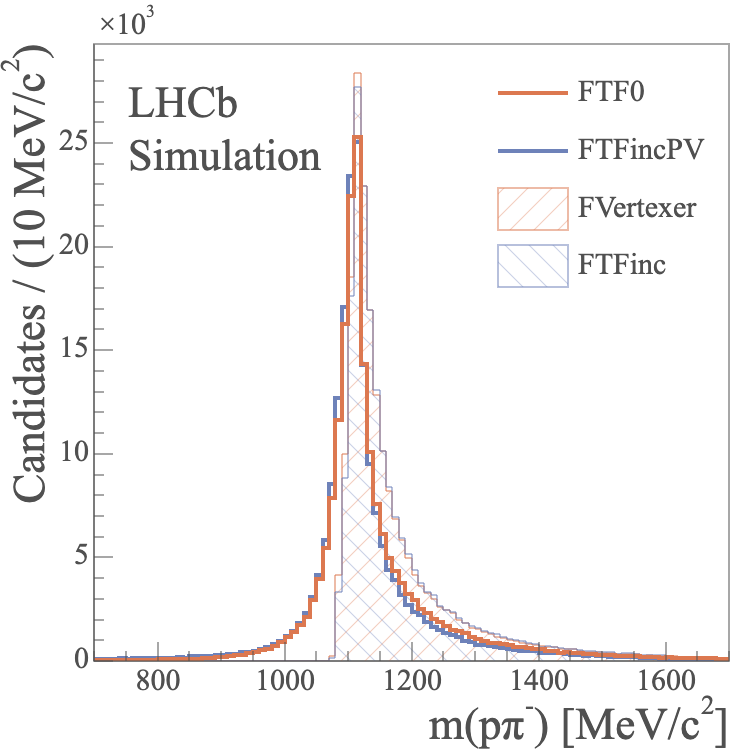}} &
{\includegraphics[width=0.245\textwidth]{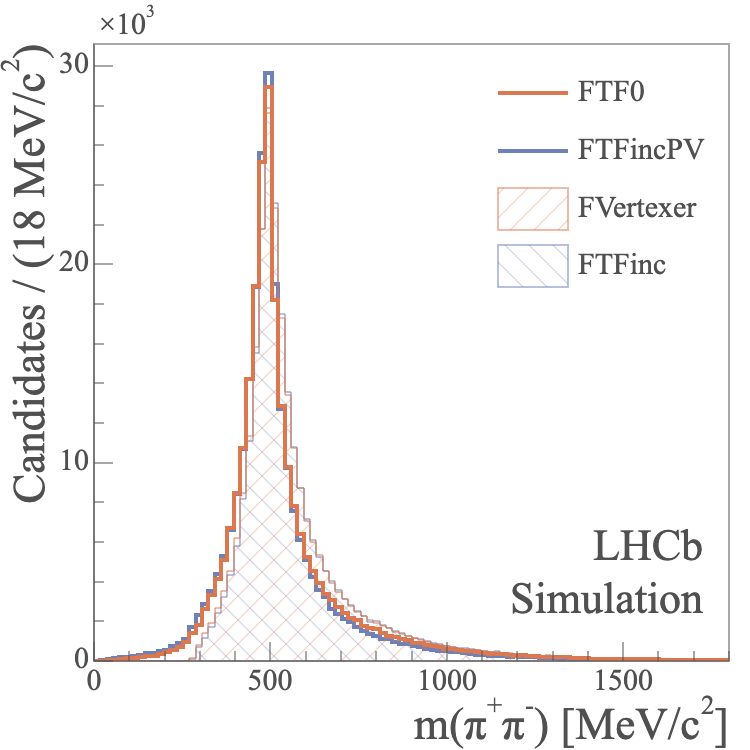}} \nonumber
\end{align}
\caption{invariant mass distributions of (left) \Lz and (right) \KS candidates from TT pairs reconstructed in inclusive mode using \ftf without and with the PV constraint (`inc' and `incPV' labels), compared with \fvertexer and with exclusive \ftf reconstruction without PV or mass constraints (label 0).
}
\label{fig:FTF_LLPs}
\end{figure}

The long-lived hadrons can also be reconstructed on their own (inclusive mode) using \ftf, both without and with the PV constraint. The corresponding invariant mass distributions are shown in Fig.~\ref{fig:FTF_LLPs}, where they are compared with those obtained using \fvertexer (also shown in Fig.~\ref{fig:FTF_LLPs}) and with \ftf when the hadrons are reconstructed as part of the full \bquark-hadron decay chains (exclusive mode), without PV or mass constraints. The results from inclusive \ftf without PV constraint and from \fvertexer are similar, since in this case the measurement and exact constraints are identical. Conversely, the inclusive \ftf with the PV measurement constraint naturally enforces the exact internal constraint that aligns the long-lived hadron momentum with its flight direction, Eq.~(\ref{eq:origin_constr}). Analogously, in the exclusive \ftf fit the role of the PV is effectively played by the \jpsi vertex position through the exact internal constraint given by~(\ref{eq:decay_vertex_constr}). These constraints cause the \Lz mass distribution to extend below the proton-pion threshold, yielding a more symmetric mass shape. If the PV constraint is additionally applied in the exclusive \ftf case (i.e. to the \bquark hadron), the mass resolution of the long-lived hadron improves significantly (not shown).

\begin{figure}[!htb]
\centering
{\includegraphics[width=0.45\textwidth]{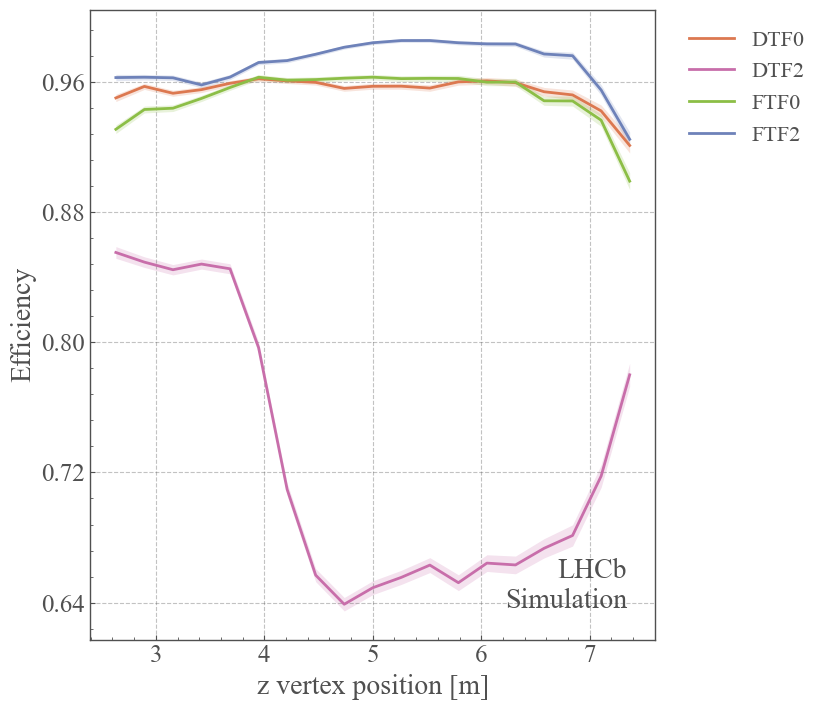}} 
{\includegraphics[width=0.45\textwidth]{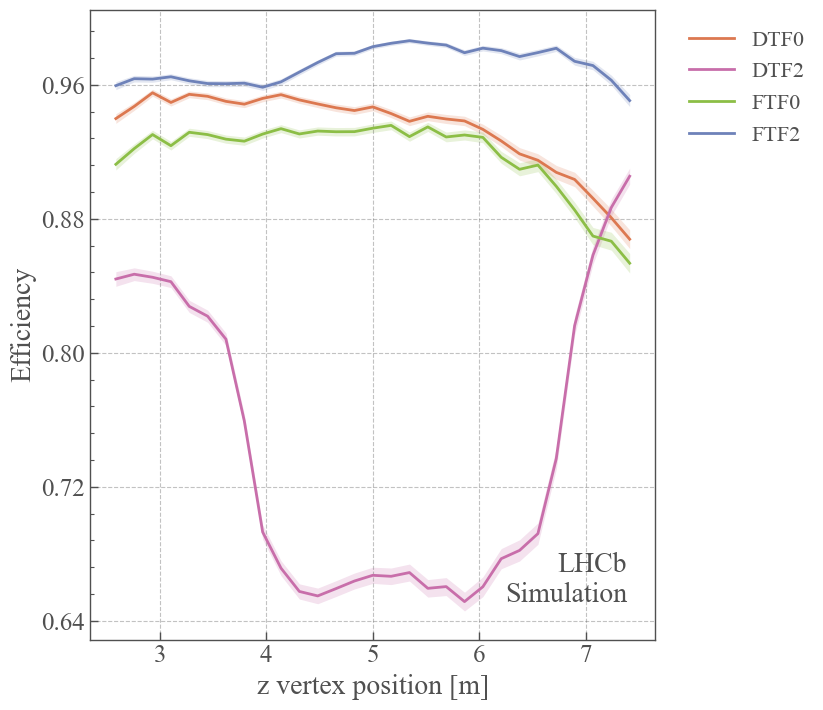}} 
\caption{Reconstruction efficiency of (top) \Lb and (bottom) \Bd signal candidates as a function of the true decay position along the $z$ axis of the \Lz and \KS hadrons from TT pairs, respectively, obtained with \ftf and \dtf. Results are shown for configurations with the PV constraint and either without (label 0) and with (label 2) the mass constraints applied to the \jpsi and long-lived hadrons.
The bands represent the corresponding 68.3\% CL intervals.
}
\label{fig:FTF_efficiency}
\end{figure}

Figure~\ref{fig:FTF_efficiency} shows the corresponding reconstruction efficiencies as a function of the true decay position along the $z$ axis of the long-lived hadron. The efficiency is defined as the fraction of signal candidates
for which the fit either satisfies the convergence criteria or reaches
the maximum number of iterations while retaining a numerically valid fit solution,
and that have a reconstructed \Lb (\Bd) invariant mass within the range
\mbox{$5300–6000~(5000–5600)\,\mev$}, evaluated with respect to the signal candidates after four-track reconstruction and passing the very loose selection applied to reduce combinatorial background. A substantial increase in efficiency is observed with \ftf when all mass constraints are enabled, once again demonstrating the impact of the \textit{global track model} and the optimised seeding strategy.

The timing performance of the algorithms is difficult to characterise in absolute terms, as it depends on the tuning, the complexity and constraints of the decay tree, and the chosen processing sequence. In practice, based on current applications within the \lhcb experiment, timing has not been a limiting factor.

\section{Conclusion}
\label{sec:conclusion}

This paper has presented \forwardtreefitter, a simultaneous Kalman-based decay tree reconstruction framework designed for complex decay topologies
involving long trajectory extrapolations, decays inside intense and inhomogeneous magnetic fields, and limited momentum resolution. Although developed primarily for the \lhcb experiment and motivated by the reconstruction of \bquark- and \cquark-hadron decays containing long-lived particles that decay inside the dipole magnet region, the framework remains sufficiently flexible to accommodate both the approximations commonly employed in conventional implementations of decay tree fitters and the geometries of general-purpose collider detectors with $4\pi$ coverage.

The main methodological innovation lies in the treatment of charged-particle trajectories. Rather than transporting track measurements and covariance matrices to the decay vertex, \ftf\ keeps the measurements fixed at their reference surfaces and evaluates the measurement residuals by propagating the fitted track states through the magnetic field. The reference implementation employs a fully nonlinear trajectory model, evaluated numerically by Runge--Kutta integration through the magnetic-field map, while only the exact constraints are iteratively relinearised around the updated solution. Together with dedicated seeding strategies and flexible handling of constraints, this provides a robust and numerically stable framework for nonlinear and tightly constrained decay topologies.

Benchmark studies demonstrate significant improvements in reconstruction efficiency, vertex position resolution and invariant mass resolution with respect to the standard \lhcb reconstruction tools for highly displaced decays occurring in the magnet region, particularly when invariant mass constraints are imposed. 
These results demonstrate the importance of evaluating the track residuals at their measurement surfaces by propagating the fitted particle trajectories through the magnetic field using a fully
nonlinear numerical description.
At the same time, the studies show that the irreducible population of events with large non-Gaussian $z$-vertex residuals, observed in \Lz\ decays and to a lesser extent in \KS\ decays, arises
from the intrinsic information-limited reconstruction regime associated with nearly collinear daughter tracks in low-$Q$ two-body decays.

\begin{acknowledgments}

We express our gratitude to our colleagues of the \lhcb collaboration, in particular the
Real Time Analysis (RTA) group, for their support and encouragement while developing this framework.
We especially thank V. Gligorov and M. Van Veghel for their help with the integration in the \lhcb software environment and the careful review the manuscript, 
the \lhcb computing and simulation teams for producing the simulated samples used to benchmark the performance of the algorithms,
and A. Fernandez Casani for his local computing support.
We also acknowledge the use of OpenAI's ChatGPT as an editorial assistant for reviewing and improving the clarity, structure, and language of the manuscript. The authors reviewed and take full responsibility for all scientific content, analyses, interpretations, and conclusions presented in this work.
We acknowledge support from
MICINN (PID2022-139842NB-C22 project) and GVA (CIPROM/2022/36 project) (Spain),
INFN (Italy), and the ERC Consolidator Grant SELDOM G.A. 771642.

\end{acknowledgments}


\bibliography{biblio}

\appendix
\section{Intrinsic limitations for low-$Q$ two-body decays}
\label{sec:performance_lowQ}

The reconstruction of highly displaced two-body decays in the targeted experimental scenario becomes
particularly challenging when the available phase space is small. This
is especially relevant for $\Lambda\to p\pi^-$ decays, for which the
low $Q$ value can produce nearly collinear daughters in the laboratory
frame. The simulated $\Lambda\to p\pi^-$ and $K^0_S\to\pi^+\pi^-$ decays
show two characteristic features in the reconstructed $z$-vertex
residual, as observed in Fig.~\ref{fig:FTF_zresol}: long and asymmetric non-Gaussian tails, and more localised and broad structures far away from zero. Both depend strongly on the decay position and become more pronounced for nearly collinear configurations. To characterise these
effects, we operationally define \emph{ghost} events as those satisfying
an absolute $z$-vertex pull reconstructed with both mass and PV constraints exceeding \eg 10 units.
This terminology is retained from earlier studies in which such events
were associated with two possible crossings of the daughter
trajectories~\cite{LHCb:2022kwc}. As discussed below, this interpretation is too
restrictive.

The long and asymmetric tails are already visible in the DOCA and standalone vertex
reconstructions, see Fig.~\ref{fig:zVertex}.
Moreover, the two features persist after several variations of the \ftf procedure, in particular when using an explicit two-body parameterisation with the
same number of degrees of freedom but a minimal set of kinematic
parameters, with different seeding and convergence
strategies, and a second-order Kalman treatment of the exact invariant
mass constraint. Notably, the same behaviour is reproduced
in a fast simulation based on unbiased Gaussian fluctuations of the
track parameters with realistic covariances. In this simplified
setting, the effect is strongly reduced either by improving the track
resolution or by artificially increasing the mass of the long-lived
particle, and hence the available phase space of the decay. These observations
indicate that the effects arise from the limited information carried
by the reconstructed tracks in the low-$Q$ regime, rather than from a
particular implementation of the fit.

Both features are correlated with the projected opening of
the daughter trajectories in the $yz$, non-bending plane. In
particular, they can be ordered by the reconstructed \emph{opening significance}
$S_{yz} = |\Delta t_y|/\sigma_{\Delta t_y}$,
where $\Delta t_y$ is the difference between the daughter trajectory slopes in the $yz$ plane and $\sigma_{\Delta t_y}$ its uncertainty.
The ghost fraction increases rapidly at small $S_{yz}$ and is
considerably larger for $\Lambda$ than for $K^0_S$ decays, consistent
with the smaller $Q$ value of the former. A requirement on $S_{yz}$ reduces both the long tails and the broad structures.

The long and asymmetric tails are a consequence of the nonlinear nature of the vertex estimator combined with the limited tracking resolution~\cite{James:2006,Hinkley:1969}. Indeed, for two locally straight trajectories and restricting to a plane for simplicity (the non-bending), the longitudinal vertex position estimator is \mbox{$z\sim \Delta y/\Delta t_y$},
and, if the uncertainty on the slope difference dominates,
\begin{eqnarray} 
\label{eq:sigmaz}
  \sigma_z & \sim & \frac{|\Delta y|}{(\Delta t_y)^2}\sigma_{\Delta t_y}.
\end{eqnarray} 
The $z$-vertex position therefore becomes increasingly poorly
determined as the trajectories approach collinearity. Moreover, the estimator can transform unbiased track fluctuations into biased and asymmetric vertex residuals. 
This can be seen explicitly by writing the measured slope difference
as $\Delta t_y+\epsilon$, with $\langle\epsilon\rangle=0$,
\begin{eqnarray} 
  \frac{\Delta y}{\Delta t_y+\epsilon} & \simeq &
  \frac{\Delta y}{\Delta t_y} -\frac{\Delta y}{(\Delta t_y)^2}\epsilon
  +\frac{\Delta y}{(\Delta t_y)^3}\epsilon^2+\cdots .
\end{eqnarray} 
The first two terms give~(\ref{eq:sigmaz}) and combined with the third term we have
\begin{eqnarray} 
  \qquad \langle z-z_{\rm true}\rangle & \sim &
    \frac{\Delta y}{(\Delta t_y)^3}\sigma_{\Delta t_y}^2,\ \ \ 
\end{eqnarray} 
and hence $|\langle z_{\rm reco}-z_{\rm true}\rangle|/\sigma_z \sim 1/S_{yz}$.
Thus, nonlinear effects are strongly amplified when the opening is poorly resolved.
For decays occurring in the $3-6$\,\m region, the tails are more pronounced towards 
positive residuals, while for $6-8$\,\m they become more negative. The local
curvature of the nonlinear estimator, and hence the sign of its
second-order contribution, can change with the decay position,
naturally allowing such a reversal of the residual asymmetry.

The broad structures represent another manifestation of the same
information-limited regime, with their relative prominence and sign
evolving with the decay position. When the daughter $yz$ opening is poorly
resolved, two possible crossings can occur in the $xz$ (bending)
plane, producing a characteristic displacement of the reconstructed
vertex. For more upstream decays, the alternative crossing tends to
lie downstream of the true decay position, thus the reconstruction of
the wrong crossing preferentially produces a positive residual.
Conversely, for more downstream decays, the alternative crossing tends
to lie upstream, preferentially producing a negative residual.
Importantly, two geometrical crossings do not necessarily result in two
distinct minima of the reconstructed fit. This has been studied by
scanning the initial $z$ seed and comparing the resulting minima, 
and explicitly scanning the global $\chi^2$ as a function of $z$.
Neither selecting events with a single minimum nor, for events with
several minima, requiring increasingly large $\Delta\chi^2$ between
the best and second-best solutions removes the structures nor the
tails. Thus, the number and relative $\chi^2$ of reconstructed minima
do not provide an effective discriminator.

Further support for these interpretations is obtained with the fast simulation by artificially reducing the track-parameter uncertainties by roughly one and two orders of magnitude. The ghost fraction decreases markedly when improving the resolution, while its dependence on small
$yz$ opening angles and its significance remains. However, a requirement on 
$S_{yz}$ retains a substantially larger fraction of correctly reconstructed events than at nominal resolution.

In summary, the two observed features in the $z$-vertex resolution are
manifestations of an intrinsic information-limited behaviour in nearly
collinear low-$Q$ decays. Poorly resolved angular
information weakens the longitudinal vertex determination: its
nonlinear propagation produces the non-Gaussian and asymmetric tails,
while the same loss of information enhances the ambiguity between
geometrical crossings inducing broad structures.

\end{document}